\documentclass{article}

\usepackage{arxiv}

\usepackage{arxiv}

\usepackage{amsmath,amstext,amsthm, amssymb,amsfonts,graphics,graphicx}
\usepackage{multicol,multirow}
\usepackage{tabularx,booktabs}
\usepackage{boites,enumerate}
\usepackage{threeparttable}

\usepackage{xcolor}

\usepackage{natbib}

\usepackage{enumitem}

\usepackage{times}
\usepackage{bm}

\usepackage{appendix} 
\usepackage{lipsum}

\usepackage{algorithm}

\newtheorem*{remark}{Remark}
\usepackage{pdflscape} 
\usepackage{listings}

\definecolor{codegreen}{rgb}{0,0.6,0}
\definecolor{codegray}{rgb}{0.5,0.5,0.5}
\definecolor{codepurple}{rgb}{0.58,0,0.82}
\definecolor{backcolour}{rgb}{0.95,0.95,0.92}

\lstdefinestyle{mystyle}{
  backgroundcolor=\color{backcolour},   commentstyle=\color{codegreen},
  keywordstyle=\color{magenta},
  numberstyle=\tiny\color{codegray},
  stringstyle=\color{codepurple},
  basicstyle=\ttfamily\footnotesize,
  breakatwhitespace=false,         
  breaklines=true,                 
  captionpos=b,                    
  keepspaces=true,                 
  numbers=left,                    
  numbersep=5pt,                  
  showspaces=false,                
  showstringspaces=false,
  showtabs=false,                  
  tabsize=2
}

\title{A Nonparametric Bayesian Local-Global Model for Enhanced Adverse Event Signal Detection in Spontaneous Reporting System Data}

\author{
 Xin-Wei Huang \\
  Department of Biostatistics\\
  University at Buffalo\\
  \texttt{xinweihuangstat@gmail.com} \\
   \And
 Saptarshi Chakraborty \\
  Department of Biostatistics\\
  University at Buffalo\\
  \texttt{chakrab2@buffalo.edu} 
}

\begin{document}

\maketitle

\begin{abstract}
    Spontaneous reporting system databases are key resources for post-marketing surveillance, providing real-world evidence (RWE) on the adverse events (AEs) of regulated drugs or other medical products. Various statistical methods have been proposed for AE signal detection in these databases, flagging drug-specific AEs with disproportionately high observed counts compared to expected counts under independence. However, signal detection remains challenging for rare AEs or newer drugs, which receive small observed and expected counts and thus suffer from reduced statistical power. Principled information sharing on signal strengths across drugs/AEs is crucial in such cases to enhance signal detection. However, existing methods typically ignore complex between-drug associations on AE signal strengths, limiting their ability to detect signals. We propose novel local-global mixture Dirichlet process (DP) prior-based nonparametric Bayesian models to capture these associations, enabling principled information sharing between drugs while balancing flexibility and shrinkage for each drug, thereby enhancing statistical power. We develop efficient Markov chain Monte Carlo algorithms for implementation and employ a false discovery rate (FDR)-controlled, false negative rate (FNR)-optimized hypothesis testing framework for AE signal detection. Extensive simulations demonstrate our methods' superior sensitivity---often surpassing existing approaches by a twofold or greater margin---while strictly controlling the FDR. An application to FDA FAERS data on statin drugs further highlights our methods’ effectiveness in real-world AE signal detection. Software implementing our methods is provided as supplementary material.
\end{abstract}

\section{Introduction}
\label{sec:intro}

Adverse events (AEs) related to drugs and other medical products have a profound impact on public health and have received heightened attention over the past decades. While clinical trials remain the gold standard for assessing medical product safety before approval, they have limitations due to factors such as small sample sizes, ethical considerations, restricted timelines, etc., and the critical need for postmarket monitoring is increasingly recognized. For example, severe AEs such as liver injury from Ketoconazole \cite{lavertu2021new} were identified only after these drugs were approved for the U.S. market. Outside the U.S., congenital malformations caused by thalidomide were identified only after their introduction in European and other markets. To complement clinical trial findings with real-world evidence (RWE), continuous post-marketing surveillance---a cornerstone of contemporary pharmacovigilance---has proven essential. Accordingly, several Spontaneous Reporting System (SRS) databases have been established globally, including VigiBase (World Health Organization), and FAERS and VAERS (U.S. Food and Drug Administration). These databases collect and manage AE data for regulated medical products, such as drugs, medical devices, vaccines, and therapeutic biologics. For exposition, we focus only on drugs hereinafter, though analogous considerations apply to other medical products monitored in SRS databases. 

The SRS databases serve as key resources for assessing real-world evidence in postmarket drug safety by cataloging a large number of case safety reports, each listing the specific drugs taken by a user along with AEs encountered. For example, the FAERS database has compiled over $29$ million reports by 2024. These databases are inherently observational, and the data curated in them suffer from several statistical challenges, including selection bias, measurement error, confounding, missing data, and absence of information on non-events, non-users, and general health characteristics of users, among other factors \cite{markatou2014pattern}, precluding formal causal inference.  In SRS data mining, the focus instead is on detecting \textit{signals}--- identifying drug-AE pairs whose \textit{observed counts} are disproportionately higher than their \textit{null expected} counts, the theoretical baseline counts for the pairs under the assumption that the drug and the AE are independent. Then signal detection entails first quantifying the \textit{signal strength}, typically in terms of some parameterization of the ratio of the observed counts to the null expected counts (or some related quantities) for each drug-AE pair, and then identifying pairs that exhibit sufficiently large signal strengths. 

An important consideration in this context is the derivation of these null expected counts for drug-AE pairs. Unlike traditional contingency table-based case-control data analysis, a key challenge here is the absence of information on non-events: SRS databases record only \textit{reported} drugs and AEs and provide no information on the total number of individuals taking the drugs who did not experience or report a specific AE. This absence means that the \textit{true denominators}---the actual counts of patients exposed to each drug---are unavailable, precluding direct calculation of absolute risks or incidence rates as in traditional epidemiological designs. Instead, SRS data mining methods rely on \textit{statistical comparators} (all other reports in the database not involving the drug or AE in question) to approximate marginal total report counts for each drug and AE. These approximated margins are then used to estimate the null expected counts for the drug-AE pairs. Despite these limitations, meaningful safety signals can be detected from these databases when appropriate statistical methods are applied, as demonstrated by a growing body of research \cite{gravel2023considerations, kontsioti2023exploring, zhang2018three, markatou2014pattern}.

For principled and accurate signal detection from SRS databases, numerous statistical data mining methods have been developed over the past decades. Examples include the proportional reporting ratio (PRR) \cite{evans2001use}, the reporting odds ratio (ROR) \cite{rothman2004reporting}, formal frequentist methods such as the likelihood ratio test (LRT) and its variants \cite{huang2011likelihood, chakraborty2022use, ding2020evaluation}, and Bayesian methods such as the Gamma Poisson Shrinker (GPS) \cite{dumouchel1999bayesian}, Bayesian confidence propagation neural network (BCPNN) \cite{bate1998bayesian}, and Dirichlet process (DP)  Poisson mixture model Hu et al.\cite{hu2015signal}. Some approaches---e.g., PRR and ROR---employ ad-hoc thresholds on empirical signal strength estimates, while the more formal methods---e.g., LRT and Bayesian approaches---employ rigorous probabilistic tools for statistically guaranteed signal detection, including controlled type I errors and false discovery rates (FDR). Despite these advances, signal detection for rare AEs and newer drugs remains challenging due to their small (marginal total) sample sizes, which leads to small observed and expected report counts in individual drugs, reducing the statistical power to detect the underlying signals. Yet, a large and growing body of research\cite{poluzzi2012data, sardella2018pharmacovigilance, raschi2021value} has documented the critical need for reliable early signal detection precisely under these settings involving rare AEs and newer drugs.
 
In this paper, we argue that an appropriately formulated Bayesian framework can yield important insights for pharmacovigilance under these challenging settings. Bayesian SRS data mining methods enable principled inference and signal detection by modeling drug-AE signal strengths as parameters with suitable priors and utilizing their posterior distributions. Prior specification, however, is crucial for optimal performance.  An overly restrictive prior---e.g., a single, common gamma prior for all drug-AE signal strength parameters---may cause excessive shrinkage, pulling estimates toward a single common value\cite{hu2015signal}. Conversely, an overly flexible prior---e.g., independent flat priors for all parameters---may inadequately regularize their estimation, resulting in overfitting and noisy estimates. State-of-the-art Bayesian data mining approaches balance flexibility and regularization using semi/nonparametric shrinkage priors, such as a two-component gamma mixture prior (the GPS model\cite{dumouchel1999bayesian}) or a DP mixture prior with a base gamma distribution\cite{hu2015signal}, for AE signal strengths \textit{separately for each drug}. These approaches stabilize signal detection; however, they still impose restrictive assumptions, treating drugs in question either entirely independently or as a single homogeneous group.  

\begin{figure}
\centering
\includegraphics[width=0.80\textwidth]{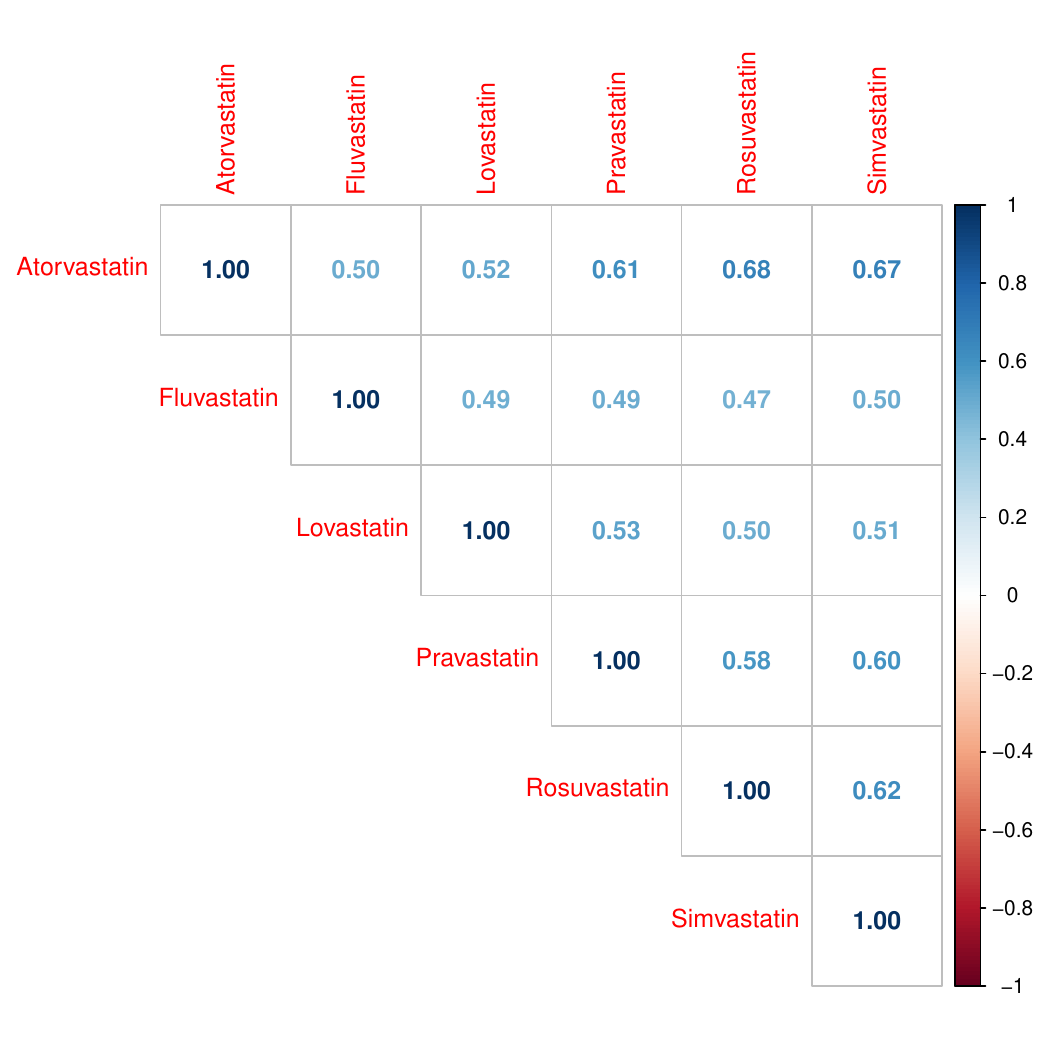} 
\caption{Kendall's $\tau$ correlation matrix for occurrence counts of 1491 commonly occurring AEs (PTs) among six statin drugs in the FDA FAERS data (2014 Q1-2020 Q4).}
\label{fig:corplot}
\end{figure}

Real-world SRS databases, however, record drugs with nuanced similarities in their AEs. For instance, various statin drugs, used to lower blood cholesterol and cardiovascular risk, share chemical compositions and function, leading to common AEs like rhabdomyolysis \cite{montastruc2023rhabdomyolysis}. Yet, due to small differences in drug composition and function they may cause distinct AEs or have variations in AE severity; e.g., myoglobinuria is typically associated with Atorvastatin but not Lovastatin \cite{seneff2016statins}.  Figure~\ref{fig:corplot} shows the pairwise Kendall's $\tau$ correlation coefficients between six statin drugs with respect to the occurrence counts of 1491 AEs as cataloged in the FAERS database (Q3 2014-Q4 2024; see Section~\ref{sec:simulation} for details on the dataset). These correlations reveal strong yet nuanced between-drug associations: they are consistently above $0.40$ with some as high as $0.68$, ruling out independence, yet many are noticeably around or below $0.50$, arguing against treating all statins as a single homogeneous group. These empirical findings illustrate a key challenge in pharmacovigilance: drugs often exhibit correlated but non-identical AE profiles, making assumptions of either independence or identicalness inappropriate and motivating flexible modeling of between-drug dependencies in SRS data mining.

There are several prominent case studies that reinforce this challenge. For example, the statin cerivastatin was withdrawn from the U.S. market {\cite{staffa2002cerivastatin}} within four years of its introduction due to a heightened risk of fatal rhabdomyolysis; this early detection was facilitated by prior knowledge that rhabdomyolysis was a recognized AE of atorvastatin and simvastatin, which had been marketed earlier.  Conversely, increased cardiovascular risk first became evident with the COX-2 inhibitor rofecoxib, which was consequently removed from the market{\cite{topol2004failing}}. Similar concerns subsequently emerged for the related COX-2 inhibitor celecoxib after the withdrawal of rofecoxib{\cite{solomon2005cardiovascular}}, until a large 10-year randomized trial (PRECISION) enrolling approximately 24,000 patients ultimately established the cardiovascular risk of celecoxib to be comparable to that of traditional nonsteroidal anti-inflammatory drugs (NSAIDs) such as ibuprofen and naproxen{\cite{nissen2016cardiovascular}}. Together, these examples illustrate that safety signals frequently arise at the drug-class level owing to between-drug similarities, while also highlighting that these associations are typically partial rather than identical across drug-specific AE profiles. Flexible acknowledgment of these partial associations is therefore necessary for effective signal detection.

Yet, existing SRS data mining approaches, both Bayesian and frequentist, rarely account for such nuanced between-drug associations. Instead, they often assume either independence of drugs  (the original LRT, BCPNN, GPS, and DP Poisson-Gamma) or a single group for all drugs (extended LRT of Huang et al.\cite{huang2011likelihood}). These restrictive assumptions limit the statistical power and \textit{sensitivity} for detecting signals, particularly for rare AEs and newer drugs with small observed and expected counts. Indeed, our extensive simulations (Section~\ref{sec:simulation}) show that existing methods suffer from poor sensitivity (after controlling FDR) when detecting signals from datasets containing multiple but weak \textit{true} signals, rare AEs, and newer drugs.   

We address this critical gap with a novel \textit{local-global mixture DP}-based Bayesian method that balances flexibility and shrinkage in AE signal strengths for each drug while \textit{enabling data-adaptive information sharing across drugs with respect to their AE patterns}, enhancing statistical power for signal discovery. In this framework, a local DP models distinct AE signal strength patterns specific to each drug while a global process defines a common pattern across drugs; the final AE signal strengths for a drug are obtained from a probabilistic mixture of these local and global processes. By appropriately controlling the local-global contributions, our approach can flexibly capture complex between-drug associations in SRS data.

For full Bayesian inference from our model, we employ weakly informative priors customized for SRS data mining on various DP hyperparameters. This eliminates the need for any additional tuning and acknowledges all modeling uncertainty, including hyperparameters. Subsequently, we develop efficient Markov chain Monte Carlo (MCMC) algorithms for posterior computation. Signal detection is performed using a rigorous hypothesis testing framework that controls the false discovery rate (FDR) while minimizing the false negative rate (FNR), both estimated via MCMC-computed posterior probabilities. Our default implementation supports a wide range of data-generating settings for robust signal detection. Additional inferential goals, such as identifying structural zero (physically or biologically impossible drug-AE pairs), can be seamlessly incorporated through additional model layers without compromising signal detection accuracy, which we also discuss. 

Our key contributions are as follows. First, we propose local-global mixture DP models for SRS signal detection (Section~\ref{sec:DP new}). Local-global priors are widely used in contemporary statistics (see Bhadra et al.\cite{bhadra2019lasso} and references cited therein) and have been employed in DP modeling of grouped data, most prominently through hierarchical DP models \cite{teh2004sharing}. Such models require handling complex combinatorial partition structures---often represented via Chinese restaurant franchise-style distributions \cite{teh2004sharing}---making implementation challenging, particularly for sizable datasets. In contrast, our framework adopts a more transparent two-component mixture of DPs \cite{antoniak1974mixtures} for local-global pattern decomposition. This enables our framework to achieve efficient MCMC implementation using standard distributions while retaining the ability to capture both shared global structure and drug-specific local variations. To our knowledge, the use of local-global DP models for principled SRS data mining is novel. We develop efficient MCMC algorithms for model implementation (Section~\ref{sec:full-bayes-infer} and Supplement~\ref{ssec:computational_details}; software in the Supplement) and leverage FDR-controlled, FNR-optimized hypothesis tests for signal detection (Section~\ref{sec:signal-detect}). Second, we provide an intuitive overview of DP-based SRS data mining models and guidance on prior specification for hyperparameters to aid broader understanding (Section~\ref{sec:DP revisit}). Third, through extensive simulations (Section~\ref{sec:simulation}), we show that our methods achieve comparable or superior sensitivity to existing approaches (reviewed in Section~\ref{sec:competing_methods}) when only a few strong signals are present across common AEs but yield substantial sensitivity gains---often twofold or more---when many weak signals exist among rare AEs and newer drugs, all while maintaining strong FDR and type I error control. Finally, our analysis of a real-world statin dataset (Section~{\ref{sec:data analysis}}) from the FDA FAERS database demonstrates our method's ability to effectively detect AE signals in practice from large-scale SRS databases, particularly for rarer AEs and newer drugs, that are often missed by existing approaches.

\section{SRS DATA, OUR NOTATION, AND PROBABILITY MODELS}
\label{sec:notation-prob-models}

An SRS database stores a large number of individual case safety reports, each documenting the list of drugs a user was exposed to and the AEs encountered. These reports can be summarized as a high-dimensional contingency table cataloging the occurrences of all AEs noted in the database across the rows and various drugs across the columns.  Further, let $n_{ij}$ be the total number/count of reported cases for drug-$j$ and AE-$i$; $i = 1, \dots, I$ and $j = 1, \dots, J$ constituting the contingency table $((n_{ij}))$. Define  $n_{i\bullet} = \sum_{j=1}^{J} n_{ij}$, $n_{\bullet j} = \sum_{i=1}^{I} n_{ij}$, and $n_{\bullet \bullet} = \sum_{i=1}^{I} \sum_{j=1}^{J} n_{ij}$. As noted in the Introduction, the true total counts of users of a drug or the total number of individuals suffering from a specific AE are not available in SRS data due to the lack of information on non-events and non-users. Instead, SRS data mining methods consider comparator drugs---all drugs other than the drug-$j$ in question---and comparator AEs --- all AEs other than the AE $i$ in question---cataloged in the SRS database, and compare the observed prevalence of the drug-$j$, AE-$i$ pair, which is substantially larger than other drug-AE pairs. Using the comparator drugs, and the \textit{reported margins} $n_{i \bullet}$, $n_{\bullet j}$, and $n_{\bullet \bullet}$ measuring the total number of \textit{reported} cases specific to AE-$i$, drug-$j$, and all drugs and AEs combined, respectively, they approximate the expected null baseline count $E_{ij} = n_{i\bullet} n_{\bullet j} / n_{\bullet\bullet}$, assuming independence between the drug and the AE. Of note, SRS data mining often focuses on identifying significant AEs of certain groups or subgroups of drugs, constituting only a (small, somewhat homogeneous) subset of all drugs cataloged in the database. Here, an artificial, heterogeneous group of drugs, obtained by collapsing all other drugs in the database, is often created as a `baseline' or `reference' category. This category of ``Other drugs'' is conceptually too heterogeneous to have any meaningful, unique AEs, but is still included to aid better comparison for disproportionality analysis, and a greater AE (row) total ($n_{i \bullet}$) counts for more accurate $E_{ij}$ estimation. Analogously, a heterogeneous group of 'Other AEs', obtained by collapsing all AEs outside those of interest (or with very small sample sizes), may be constructed \cite{ding2020evaluation}.    

Signal detection in SRS contingency tables typically relies on the observed relative reporting rates, $n_{ij}/E_{ij}$, or a related quantity, as a measure of \textit{signal strength}. Intuitively, if $n_{ij}/E_{ij} \gg 1$, the observed frequency for $(i, j)$ is much higher than the null expected count, suggesting a potential signal. Challenges arise for small $n_{\bullet j}$ (e.g., a new drug) and/or small $n_{i \bullet}$ (rare AE), leading to a small $E_{ij}$ and rendering $n_{ij}/E_{ij}$ to be noisy. Bayesian approaches parameterizing $n_{ij}/E_{ij}$ (or an analogous quantity) using a probability model for $n_{ij}$ can stabilize inference on signal strengths under these setups by incorporating suitable prior distributions that can facilitate principled information sharing across AEs and drugs, leading to improved signal detection. 

The Poisson model $n_{ij} \sim \text{Poisson}(\lambda_{ij} E_{ij})$ is commonly used for its analytical simplicity and computational tractability. Here the \textit{signal strength parameter} $\lambda_{ij}$ parametrizes $n_{ij}/E_{ij}$, with the pair $(i, j)$ considered a signal if $\lambda_{ij} \gg 1$ and non-signal otherwise. When the Poisson assumption is violated, particularly due to an excess of zero counts arising from structural zeros---i.e., drug-AE pairs that are physically or biologically impossible to co-occur---a zero-inflated Poisson (ZIP) model is often preferred. In this framework, $n_{ij}$ is modeled as a mixture of two components: a $\text{Poisson}(\lambda_{ij} E_{ij})$ component for drug-AE pairs that can co-occur and a point mass at $0$, denoted by $\delta_{\{0\}}$ (the Kronecker delta), for structural zeros. Importantly, the structural zero positions are typically unknown and inferred from the SRS dataset.

A Bayesian SRS data mining approach begins with such a probability model for $n_{ij}$ (e.g., Poisson or ZIP), yielding a likelihood for the signal strengths $\lambda_{ij}$ that is combined with a prior distribution to derive the posterior distribution of $\lambda_{ij}$ for inference. The marginal distribution of $n_{ij}$, after integrating over the prior, can be more flexible than the likelihood model for $n_{ij}$. Indeed, with an appropriate nonparametric mixture prior, a flexible marginal distribution accommodating ZIP-like behavior can be achieved using only a Poisson likelihood for $n_{ij}$ (see Section~\ref{sec:DP revisit}). Thus, Bayesian SRS data mining methods are less sensitive to the likelihood choice (Poisson or ZIP) when a flexible prior is used. However, if the zero-inflation probability parameters are of interest, the ZIP model remains useful. 

\section{A REVIEW OF A FEW EXISTING SRS DATA MINING METHODS} \label{sec:competing_methods}

As noted in the Introduction, numerous SRS data mining methods have been proposed in the literature over the past decades. As representatives of the state-of-the-art and competitors of our proposed approach, we focus on four prominent and commonly used SRS data mining methods---both Bayesian and frequentist---that enable (upon appropriate adjustment, as needed) strong false discovery rate controls while permitting reasonable sensitivity for AE signal detection. This section reviews these methods, highlighting the underlying modeling assumptions. 

\subsection{The Bayesian Propagation Neural Network (BCPNN) method} \label{sec:BCPNN}

The BCPNN approach (Bate et al.\cite{bate1998bayesian}) assumes that conditional on the grand total count $n_{\bullet \bullet}$, the cell-specific counts $n_{ij}$, the marginal row totals $n_{i \bullet}$, and the marginal column totals  $n_{\bullet j}$ each distributed as binomial-beta:
(a) $n_{ij} \mid p_{ij} \sim \text{Binomial}(n_{\bullet \bullet}, p_{ij}), \ p_{ij} \sim \text{Beta}(\alpha_{ij}, \beta_{ij})$, 
(b) $n_{i\bullet} \mid p_{i\bullet} \sim \text{Binomial}(n_{\bullet \bullet}, p_{i\bullet}), \allowbreak \ p_{i\bullet} \sim \text{Beta}(\alpha_{i\bullet}, \beta_{i\bullet})$, 
(c) $n_{\bullet j} \mid p_{\bullet j} \sim \text{Binomial}(n_{\bullet \bullet}, p_{\bullet j}), \allowbreak \ p_{\bullet j} \sim \text{Beta}(\alpha_{\bullet j}, \beta_{\bullet j})$, where $p_{ij}$ is the cell-specific reporting rate, and $p_{i\bullet}$ and $p_{\bullet j}$ are the marginal row and column reporting rates. We note that the binomial model assumptions for $n_{ij}$, $n_{i \bullet}$ and $n_{\bullet j}$ can be understood as consequences of an independent Poisson model for $n_{ij}$ with certain choices of $p_{ij}$, $p_{i \bullet}$, and $p_{\bullet j}$. Bate et al.\cite{bate1998bayesian} consider the information content $\text{IC}_{ij} = \log_2 \left[p_{ij} / (p_{i \bullet} p_{\bullet j})\right]$ as a measure of signal strength for the drug-$j$ AE-$i$  pair, with a higher IC indicating a stronger association. The authors suggest an empirical Bayes estimation of these quantities using $\hat\alpha_{ij} = \hat\alpha_{i\bullet} = \hat\beta_{i\bullet} = \hat\alpha_{\bullet j} = \hat\beta_{\bullet j} = 1$ and $\hat\beta_{ij} = 1/(E(p_{i\bullet} \mid n_{i\bullet}) E(p_{\bullet j} \mid n_{\bullet j})) - 1$, with the denominator of $\hat \beta_{ij}$ being computed using the posterior distributions $p_{i\bullet} \mid n_{i\bullet} \sim \text{Beta}(\hat\alpha_{i\bullet} + n_{i\bullet}, \hat\beta_{i\bullet} + n_{\bullet\bullet} - n_{i\bullet})$ and $p_{\bullet j} \mid n_{\bullet j} \sim \text{Beta}(\hat\alpha_{\bullet j} + n_{\bullet j}, \hat \beta_{\bullet j} + n_{\bullet\bullet} - n_{\bullet j})$. The consequent (empirical Bayes) posterior distribution of $\text{IC}_{ij}$ can be approximated using Monte Carlo draws or asymptotic normality. A signal is detected at the drug-$j$ and AE-$i$ pair if the 2.5th posterior percentile for $\text{IC}_{ij}$ exceeds $0$. While this fixed threshold of $0$ is not necessarily guaranteed to control false discoveries, methods to generate data-driven thresholds ensuring a predefined FDR level (e.g., $0.05$) have been proposed.\cite{ahmed2010false}

\subsection{The Gamma Poisson Shrinker (GPS) model} \label{sec:GPS}

The GPS method (Dumouchel \cite{dumouchel1999bayesian}) employs an empirical Bayes approach for stabilized signal detection within a Poisson model-based framework for the report counts $n_{ij}$ assuming a Poisson likelihood. Specifically, one assumes $n_{ij} \sim$ Poisson$(E_{ij} \lambda_{ij})$, with  $\lambda_{ij}$ parametrizing the signal strength for the drug-$j$ and AE-$i$ pair. A two-component gamma mixture prior is placed on $\lambda_{ij}$:
\begin{gather*}
   \lambda_{ij} \sim \kappa \text{Gamma}(\alpha_1, \beta_1) + (1 - \kappa) \text{Gamma}(\alpha_2, \beta_2),
\end{gather*}
where $\kappa \in [0,1]$ is the mixing weight hyperparameter, and $\alpha_1, \beta_1 > 0$, and $\alpha_2, \beta_2 > 0$ are mixture component-specific gamma hyperparemeters. Here Gamma($\alpha$, $\beta$) denotes a probability distribution with density $f_{\text{gamma}}(y \mid \alpha, \beta) = \frac{1}{\Gamma(\alpha)}\beta^\alpha y^{\alpha - 1} e^{-\beta y}$, $y>0$; $\alpha, \beta > 0$. Intuitively, the two components of the mixture are expected to represent clusters of $\lambda_{ij}$ corresponding to large values (``signals'') and small values (``non-signals").  The posterior distribution of $\lambda_{ij}$  given $n_{ij}$, $E_{ij}$, and the hyperparameters is:
\[
   \lambda_{ij} \mid n_{ij}, E_{ij} \sim \kappa_{ij} \text{Gamma}(\hat{\alpha_1} + n_{ij}, \hat{\beta_1} + E_{ij}) + (1 - \kappa_{ij}) \text{Gamma}(\hat{\alpha_2} + n_{ij}, \hat{\beta_2} + E_{ij})
\]
where $\kappa_{ij}=\kappa f_{\text{Gamma}}(\lambda_{ij} \mid \alpha_1, \beta_1) / [\kappa f_{\text{Gamma}}(\lambda_{ij} \mid \alpha_1, \beta_1) + (1-\kappa)f_{\text{Gamma}}(\lambda_{ij} \mid \alpha_2, \beta_2)]$. Under an empirical Bayes approach, the point estimates $\hat \kappa$, $\hat \alpha_1$, $\hat \beta_1$, $\hat \alpha_2$, and $\hat \beta_2$ for the hyperparameters are obtained by maximizing their marginal likelihoods given $n_{ij}$ and $E_{ij}$. The resulting posterior of $\lambda_{ij}$ conditional on these estimated hyperparameters is used for inference and signal detection. Specifically,  \cite{dumouchel2001empirical} define the empirical Bayesian score $E(\log_2(\lambda_{ij}) \mid n_{ij}, E_{ij}, \hat \kappa, \hat \alpha_1, \hat \beta_1, \hat \alpha_2, \hat \beta_2)$ as a point estimate of signal strength for the drug-$j$ and AE-$i$ pair, and also suggest determining the pair to be a signal if the 5th percentile (EB05) of the above conditional posterior for $\lambda_{ij}$ exceeds a specific threshold,  commonly 2. Similar to the BCPNN method, the deterministic threshold of $2$ may not necessarily ensure strong FDR control; however, a data-driven threshold controlling FDR to a pre-specified level (e.g., 0.05) may be obtained.\cite{ahmed2010false}

\subsection{DP Poisson method of Hu et al.\cite{hu2015signal}} \label{sec:DP_Hu}

Several extensions and generalizations of the two-component mixture prior used in the GPS method have been suggested in the literature\cite{hu2015signal,zhang2018three, tan2025flexible}. Of these, we focus on one of the most flexible generalizations as proposed in Hu et al.\cite{hu2015signal}, which employs a Dirichlet process mixture as the prior for $\lambda_{ij}$.  Hu et al.\cite{hu2015signal} consider the same Poisson likelihood $n_{ij} \sim$ Poisson($\lambda_{ij} E_{ij}$) as in GPS, but replace the two-gamma  mixture prior for $\lambda_{ij}$ by a DP (infinite) mixture prior:
\begin{equation} \label{eq:DP-local}
\lambda_{ij} \sim D_j \equiv \sum_{h=1}^{\infty} \eta_{hj} \delta_{\{\theta_{hj}\}}; \ D_j   \sim \text{DP}(\alpha_j, G(\beta_j)).
\end{equation}
Here $\delta_{\{\theta\}}$ denotes a degenerate distribution at $\theta$ and DP$(\alpha, G(\beta))$ is a Dirichlet process with concentration hyperparameter $\alpha > 0$ and base distribution $G(\beta) \equiv$ Gamma(shape = $\beta$, rate = $\beta$) with hyperparameter $\beta > 0$. The discrete mixture structure ensures that $D_j$, the prior for $\lambda_{ij}$, is almost surely discrete/categorical placing probability weights $\eta_{hj}$; $0 \leq \eta_{hj} \leq 1;\sum_{h=1}^{\infty} \eta_{hj} = 1$ on \textit{atoms}  $\theta_{hj}$. This permits \textit{clustering}, allowing multiple $\lambda_{ij}$ to share an atom $\theta_{hj}$ as a common value. The probability weights $\eta_{hj}$ are generated through the DP concentration hyperparameter $\alpha_j > 0$, and a constructive way to describe their generation is through the stick-breaking representation \cite{sethuraman1994constructive}:
\begin{equation} \label{eq:stick-breaking}
   \eta_{1j} = v_{1j}, \ \eta_{hj} = v_{hj} \prod_{t=1}^{h-1}(1 - v_{tj}) \text{ for } h \geq 2; \ v_{hj} \sim \text{independent Beta}(1, \alpha_j) \text{ for } h \geq 1.   
\end{equation}
Meanwhile, the atoms $\theta_{hj}$ are generated independently from the DP base gamma distribution: $\theta_{hj} \sim G(\beta_j)$; $h \geq 1$. 
In practice, the infinite mixture \eqref{eq:DP-local} defining $D_j$ can be well-approximated \cite{ishwaran2002exact} by a finite mixture $\sum_{h=1}^{K} \eta_{hj} \delta_{\{\theta_{hj}\}}$ with a sufficiently large number of components $K$; this finite mixture approximation provides a computationally tractable approach to implementing this model.  

Hu et al.\cite{hu2015signal} consider full Bayesian inference on $\lambda_{ij}$ by first eliciting independent vague prior distributions on the hyperparameters $\alpha_j$ (specifically a Uniform(0.2, 10) prior) and $\beta_j$ (a Uniform(0, 1) prior) and then utilizing MCMC sampling from the resulting joint posterior. Subsequently, the drug-$j$ and AE-$i$ pair is identified as a signal if the 5th posterior percentile of $\lambda_{ij}$ computed using its posterior MCMC draws exceeds a certain threshold. The authors suggest using the deterministic threshold of $2$ for this purpose; a more data-driven threshold controlling FDR at a prespecified level (e.g., $0.05$) may be obtained.\cite{ahmed2010false} 

Our proposed approach is motivated by this method, and we revisit this approach elucidating the key roles played by the model hyperparameters $\alpha_j$ and $\beta_j$, and guiding their prior specification for full Bayesian inference in Section~\ref{sec:DP revisit}.

\subsection{ZIP model based Pseudo LRT (frequentist) method} \label{sec:pvlrt}
In parallel to Bayesian developments, major advances on SRS data mining for AE signal detection have also been made using frequentist methods, notably using the machinery of the likelihood ratio test \cite{huang2011likelihood, huang2013likelihood, huang2014likelihood}. These approaches often have the advantage of built-in FDR control while ensuring high sensitivity. We consider the ZIP model for the reporting count $n_{ij} \sim \text{ZIP}(n = n_{ij} \mid \lambda_{ij} \times E_{ij}, \omega_j)$. The pseudo LRT, as proposed by Chakraborty et al.\cite{chakraborty2022use}, is a recent method designed to accommodate the zero-inflated nature of SRS data while easing computational demands. Similar to the standard LRT method, this approach first tests a global alternative hypothesis $H_{aj}: \lambda_{ij} > 1$ for at least one $i \in \{1, \dots, I\}$ against the null hypothesis $H_{0j}: \lambda_{ij} = 1$ for all $i \in \{1, \dots, I\}$ for a given drug-$j$. If $H_{aj}$ is rejected, a step-down procedure is used to conduct individual tests on $H_{aij}: \lambda_{ij} > 1$.

The ordinary LRT estimates the nuisance parameter $\omega_j$ separately for $H_{0j}$ and $H_{0j} \cup H_{aj}$, while the pseudo LRT is constructed using a global point estimate $\hat{\omega}_{j}$ of $\omega_j$ consistent under $H_{0j} \cup H_{aj}$ in both the numerator and the denominator of the likelihood ratio test statistic. Chakraborty et al consider the point estimate of $\omega_j$ is obtained by maximizing the profile likelihood $\ell_j(\omega_j)$ for this purpose, with $\hat{\lambda}_{ij} = \max\{1, n/E_{ij}\}$ as the maximum likelihood estimate of $\lambda_{ij}$. The resulting statistic is given by
\[
\text{LR}_{ij} = \exp\left[-(\hat{\lambda}_{ij} - 1) E_{ij} \right] \hat{\lambda}_{ij}^{n_{ij}}.
\]
The maximum likelihood ratio statistic for drug-$j$ is then defined as $\text{MLR}_{j} = \max_{\substack{i \in \{1, \dots, I\}}} \text{LR}_{ij}$. In the global test, a parametric bootstrap is used to approximate the p-value $P_{j} = \Pr(\text{MLR}_{j} \geq \text{MLR}_{j}^{\text{obs}})$ using (parametric bootstrap) draws for $\text{LR}_{ij}$ under $H_{0j}$ and hence $\text{MLR}_{j}$, where $\text{MLR}_{j}^{\text{obs}}$ is the observed test statistic. If the global test is rejected,  step-down p-values \cite{chakraborty2022use,huang2011likelihood} are computed for the individual tests using the \textit{same} parametric bootstrap samples for $\text{MLR}_{j}$ and using the individual observed likelihood ratio test statistic values, such that $P_{ij} = \Pr(\text{LR}_{ij} \geq \text{MLR}_{ij}^{\text{obs}})$. For multiple drugs, the same approach can be applied, defining an overall maximum likelihood ratio as $\text{MLR} = \max_{\substack{i \in \{1, \dots, I\}, j \in \{1, \dots, J\}}} \text{LR}_{ij}$. This step-down procedure employing the same global test statistic strictly controls the false discovery rate at the same level $\alpha$ of the global test performed using the MLR \cite{huang2011likelihood, chakraborty2022use}. When multiple drugs $j$ are tested simultaneously, one considers the \textit{extended} version with global maximum likelihood ratio $\max_{j \in \{1, \dots, J\}} \text{MLR}_j$.

\section{A NOVEL LOCAL-GLOBAL DP MODEL FOR SRS DATA MINING} 

We begin this section by revisiting the DP Poisson model of Hu et al. \cite{hu2015signal} (Section~\ref{sec:DP revisit}), highlighting the role of its hyperparameters in SRS signal discovery with the aid of a schematic plot (Figure~\ref{fg:DP}a) and providing guidance on  prior specification for these hyperparameters. We then introduce our novel local-global DP framework (Section~\ref{sec:DP new}), which accommodates between-drug associations and enables effective information sharing across drugs, thereby enhancing SRS data mining.

\subsection{The DP Poisson model revisited: the roles of the hyperparameters in SRS signal discovery and the effect of zero-inflation} \label{sec:DP revisit}

\begin{figure}[]
\centering
\includegraphics[width=\textwidth]{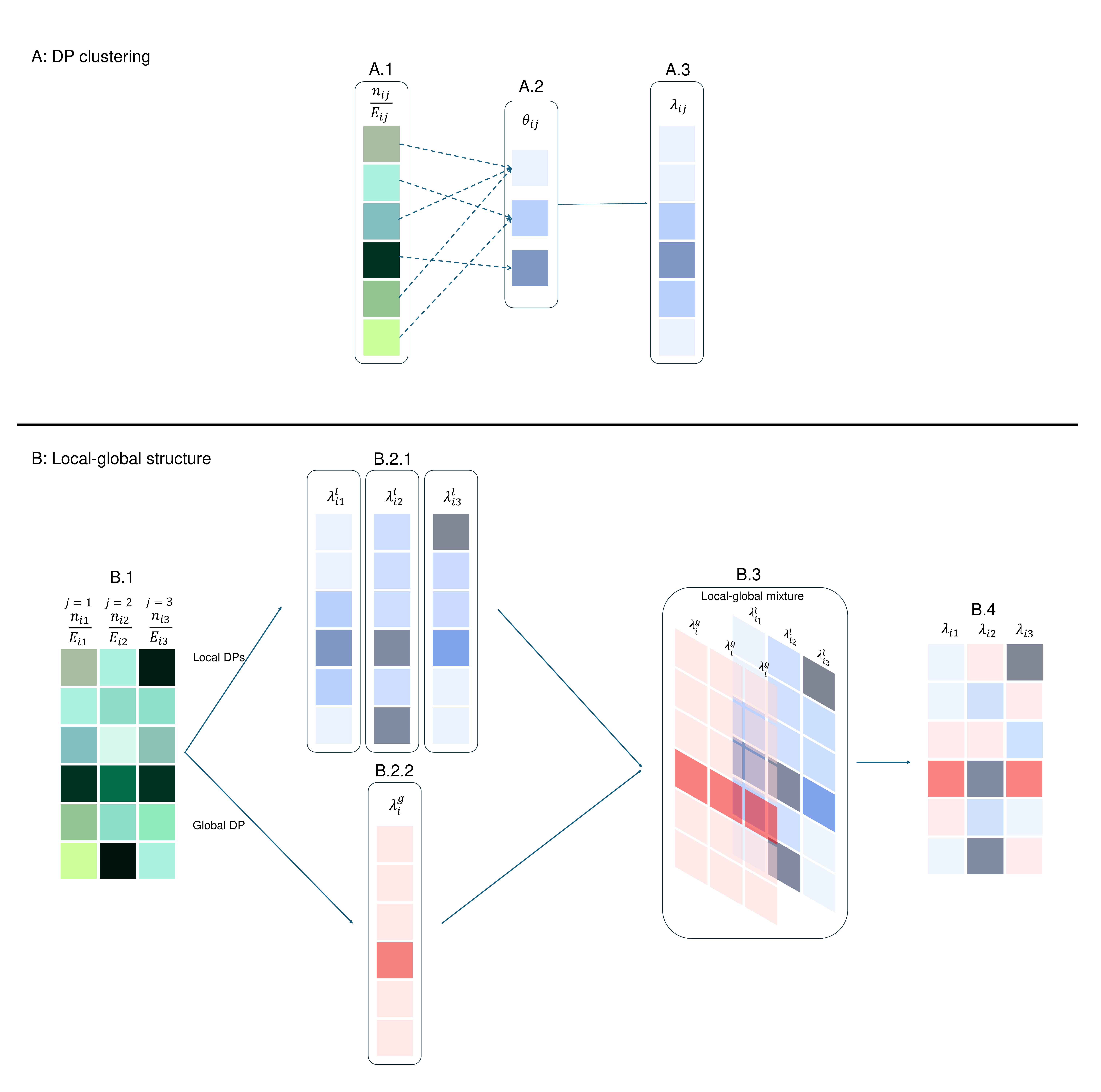} 

\caption{Schematic diagrams visualizing the mechanisms of a ``local only'' DP prior with one drug  (\textbf{Panel A}) and local-global mixture DP prior with $J = 3$ drugs (\textbf{Panel B}). Throughout, darker shades represent larger values. \textbf{Panel A:} The observed relative reporting rates $n_{ij}/E_{ij}$ for $I=6$ AEs for the drug $j$ (Panel A.1) show high variability across $i$ (different shades of green) due to randomness in observed data. The \textit{atoms} $\theta_{hj}$ (Panel A.2) of the DP prior enable clustering of $I = 6$ AE signal strengths into $K_{+} = 3$ non-empty DP clusters. This ensures information sharing: the AEs $(i = 1, i = 3, i = 5)$ in cluster 1 and separately the AEs $(i = 2, i = 6)$ in cluster 2 (cluster memberships are visualized via arrows between Panels A.1 and A.2) share their observed $(n_{ij}, E_{ij})$ information to inform their cluster-specific DP atoms (single atom per cluster). These DP atoms, together with the cluster memberships, produce the final $I=6$ AE signal strengths $\lambda_{ij}$ (Panel A.3).  \textbf{Panel B:} The noisy observed data $(n_{ij}, E_{ij})$ leading to noisy observed relative reporting rates $n_{ij}/E_{ij}$  (Panel B.1; different shades of green) are used to produce ``local'' signal strengths $\lambda_{ij}^l$ for $I = 6$ AEs and $J=3$ drugs using a separate, independent local DP prior for each drug (Panel B.2.1), and global AE signal strengths $\lambda_i^g$ for shared by all $J = 3$ drugs based on a common global DP prior (Column B.2.2). Probabilistic two-component mixtures of $\lambda_{ij}^l$ and $\lambda_{i}$ (Panel B.3) based on the local distribution indicators $z_{ij}$ produce the final realized signal strengths $\lambda_{ij}$ (Panel B.4).}
\label{fg:DP}
\end{figure}

\paragraph*{Role and prior specification for the hyperparameter $\alpha_j$} The stick-breaking representation \eqref{eq:stick-breaking} elucidates that when $\alpha_j < 1$, each $v_{hj}$, \textit{on average}, is greater than $1/2$ with smaller values of $\alpha_j$ favoring $v_{hj}$ closer to one. Consequently, $\eta_{hj}$ decreases sharply in $h$, leading the mixture \eqref{eq:DP-local}, or its finite approximation with $K$ components, to retain only a few \textit{non-empty} components with sufficiently large and positive weights $\eta_{hj}$. Under a finite mixture approximation with $K_{+} < K$ non-empty components, \eqref{eq:DP-local} models $\lambda_{1j}, \dots, \lambda_{Ij}$ using a categorical distribution with $K_{+}$ categories, placing point masses at the atoms $\theta_{1j}, \dots, \theta_{K_+ j}$ with weights $\eta_{1j}, \dots, \eta_{K_+ j}$. This fosters  \textit{clustering}, restricting $\lambda_{1j}, \dots, \lambda_{Ij}$ to at most $K_{+}$ distinct values. When $K_{+} < I$, multiple $\lambda_{ij}$ share the same cluster $h$ admitting the common value $\theta_{hj}$, facilitating their pooling for information sharing. It follows that a smaller $\alpha$ encourages a smaller $K_{+}$, resulting in fewer clusters and thus greater information sharing among the $\lambda_{ij}$, while a larger $\alpha$ induces a larger $K_{+}$, resulting in more clusters that collectively capture more hetergeneity among the $\lambda_ij$ values. Importantly, when supported by the data and induced through a (moderately) large $\alpha$, values of $K_+ >2$ are permitted in the posterior of this method. This contrasts with the GPS method of \cite{dumouchel1999bayesian}  (see Section~\ref{sec:competing_methods} for a review) which strictly categorizes $\lambda_{ij}$ into two groups, intuitively, \textit{signal} ($\lambda_{ij} \leq 1$) and \textit{non-signal} ($\lambda_{ij} \gg 1$). The DP framework, by allowing $K_{+} > 2$, enhances heterogeneity acknowledgment within both signal and non-signal groups.  Figure~\ref{fg:DP}a illustrates this with $I=6$ AEs and $K_{+}=3$ clusters. 

Because a smaller $\alpha_j$ promotes greater information sharing, a prior assigning sufficient mass to small positive values is recommended for stable estimation of $\lambda_{ij}$. Hu et al.\cite{hu2015signal} used a Uniform(0.2, 10) prior for $\alpha_j$. We suggest a more aggressive weakly informative default prior, viz., Exponential(rate = $\psi_\alpha$) with a moderately large $\psi_\alpha > 1$. This prior favors small $\alpha_j$ while allowing larger values if needed and  ensures conjugacy for efficient Gibbs sampling updates in the MCMC implementation. As a default choice, we suggest $\psi_\alpha = 3$, which places $\approx 95\%$ prior mass in $(0, 1)$. However, our simulation-based sensitivity analyses using typically sized SRS datasets indicate that model performance is largely robust to the choice of $\psi_\alpha$ when specified within a weakly informative range; see Remark~{\ref{remark:sensitivity-analysis}}.

\paragraph*{Role and prior specification for the hyperparameter $\beta_j$} The parameter $\beta_j$ controls variability among the atoms $\theta_{hj}$ drawn from the base $G(\beta_j) \equiv$ Gamma(shape = $\beta_j$, rate = $\beta_j$) distribution, thereby controlling the shrinkage in $\lambda_{ij}$. Since $G(\beta_j)$ has unit mean and variance $1/\beta_j$, larger (smaller) values of $\beta_j$ encourage all $\theta_{hj}$ to concentrate more (less) around $1$, inducing more (less) shrinkage in $\lambda_{ij}$ (around $1$). 

To balance shrinkage (induced by large $\beta_j$) with flexibility (small $\beta_j$), a prior that assigns sufficient mass to both small ($<1$) and large ($>1$) values of $\beta_j$ is preferred. Hu et al.\cite{hu2015signal} used a Uniform(0,1) prior. We suggest a \textit{half-Cauchy} prior, $\beta_j \sim$ Cauchy$^{+}(0, \psi_{\beta})$, with a moderate scale parameter $\psi_{\beta}$. We recommend $\psi_\beta = 1/2$ as a default choice, as it ensures equal prior probabilities for $[\beta_j < 1]$ and $[\beta_j \geq 1]$. However, our simulation results with typically sized SRS datasets indicate that model performance is robust to the choice of $\psi_\beta$ when specified within a weakly informative range; see Remark~{\ref{remark:sensitivity-analysis}}. The half-Cauchy prior has been recommended in the literature as a weakly informative default prior for scale parameters in Bayesian hierarchical models \cite{polsonHalfCauchyPriorGlobal2012}. (Note that $\beta_j \sim$ Cauchy$^{+}(0, \psi_{\beta})$ implies $\text{var}(\theta_{hj}) = 1/\beta_j \sim$ Cauchy$^{+}(0, 1/\psi_{\beta})$.)

\paragraph*{Effect of zero-inflation}
Zero inflation can be handled to some extent in the DP Poisson model through a small atom $\theta_{hj} \approx 0$, causing $\lambda_{ij}$ and hence Poisson($\lambda_{ij} E_{ij}$) to concentrate near zero for all AEs $i$ in cluster $h$,  effectively mimicking a degenerate-at-zero $\delta_{\{0\}}$ distribution. This yields a model similar to $n_{ij} \sim$ ZIP($\lambda_{ij} E_{ij}, \omega_{j})$, $\omega_j$ being the structural zero probability, with an analogous DP prior on $\lambda_{ij}$, producing similar posterior for $\lambda_{ij}$. Consequently, signal detection and inference on $\lambda_{ij}$ are also expected to be broadly similar in both models (Section~\ref{sec:simulation} provides numerical validation). However, some (small) differences may arise in practice, particularly when analyzing large real-world SRS datasets with complex signal patterns, due to numerical rounding errors for handling zero atoms. Furthermore, for direct inference on the structural zero positions, the Poisson approximation of $\delta_{\{0\}}$ is insufficient, and a ZIP model $n_{ij} \sim \text{ZIP}(\lambda_{ij}E_{ij}, \omega_{j})$ is needed. The same prior \eqref{eq:DP-local-global} may still be used on $\lambda_{ij}$  assuming $\lambda_{ij} = 0$ whenever the underlying cell-specific zero-inflation indicator, say $y_{ij}$, indicates a structural zero to ensure identifiability. For any drug-$j$, the posterior distributions of $y_{ij}$ and/or the column-specific zero-inflation probability $\omega_j$ provide inference on whether AE-$i$ and/or any AE is a structural zero, respectively. In practice, it is thus important to assess whether structural zeros/zero inflation are likely in the SRS dataset being analyzed, and whether inference on them is of interest, in order to guide appropriate model selection. The DP-Poisson framework offers some robustness against model misspecification due to zero inflation, but explicit ZIP modeling may be preferred when structural zeros are a primary concern.

\subsection{The local-global DP Poisson model} \label{sec:DP new}
The DP model in Section~\ref{sec:DP revisit} enables flexible and effective signal detection for each drug separately but does not account for associations between drugs in their signal strengths. To capture these associations, we introduce the concept of local and global (prior) distributions for AE signal strengths. Specifically, we assume that the realized AE signal strengths $\{\lambda_{1j}, \dots, \lambda_{Ij}\}$ for each drug $j$ is drawn from a mixture of a ``local'' distribution specific to the drug and a ``global'' distribution shared among drugs with similar compositions or functions. 

Under the same likelihood $n_{ij} \sim \text{Poisson}(\lambda_{ij}E_{ij})$, we consider the following local-global mixture of DP prior for $\lambda_{ij}$:
\begin{gather}
    \lambda_{ij} = z_{ij} \lambda_{ij}^l + (1 - z_{ij}) \lambda_i^g; \ \lambda_{ij}^l  \sim D_j; \ \lambda_{i}^g  \sim  D; \  z_{ij} \sim \text{Bernoulli}(\pi) \nonumber \\
    D_j \sim \text{DP}(\alpha_j, G(\beta_j)); \ D \sim \text{DP}(\alpha, G(\beta)). 
    \label{eq:DP-local-global}
\end{gather}

\noindent Here, the realized signal strengths $\{\lambda_{1j}, \dots, \lambda_{Ij}\}$ for a drug $j$ is generated by combining local strengths $\{\lambda_{1j}^l, \dots, \lambda_{Ij}^l\}$ drawn from a drug-$j$-specific local distribution $D_j \sim \text{DP}(\alpha_j, G(\beta_j))$ as in \eqref{eq:DP-local} and global strengths $\{\lambda_{1}^g, \dots, \lambda_{I}^g\}$ generated from a global distribution $D \sim \text{DP}(\alpha, G(\beta))$ shared by all drugs. The binary \textit{local distribution indicators} $z_{ij}$ governed by the \textit{common local proportion} $\pi$, determine the probabilistic mixing of $\lambda_{ij}^l$ and $\lambda_i^g$.  

We note that the proposed local-global mixture DP prior \eqref{eq:DP-local-global} enables information sharing in signal strengths across both AEs and drugs. Specifically, for each drug $j$, the realized signal strengths $\{\lambda_{i'j}: \text{all } i' \text{ with } z_{i'j} = 1\}$ exhibit the same $D_j$-induced clustering (see Section~\ref{sec:DP revisit}) as the local strengths $\{\lambda_{1j}^l, \dots, \lambda_{Ij}^l\}$, facilitating information sharing across AEs. Meanwhile, for any AE $i$, the signal strengths $\{\lambda_{ij'}: \text{all } j' \text{ with } z_{ij'} = 0\}$ form a cluster attaining the common global value $\lambda_i^g$, and enable information sharing across drugs (the global values $\{\lambda_1^g, \dots, \lambda_I^g\}$ themselves admitting clustering via $D$). Figure~\ref{fg:DP}B illustrates this mechanism using $I=6$ AEs and $J=3$ drugs. This bi-level clustering contrasts with the single-level (AE-only) clustering induced by the DP prior of Hu et al.\cite{hu2015signal}. As demonstrated in extensive simulations (Section~\ref{sec:simulation}), the bi-level clustering and information sharing play a critical role in stabilizing the posterior and enhancing signal detection for $\lambda_{ij}$, particularly for drug-AE pairs where $n_{i\bullet}$ is small (rare AEs) and/or $n_{\bullet j}$ is small (newer drugs), leading to small $E_{ij}$.

The drugs under consideration should be similar in composition or function to justify a single global distribution $D$. If distinct drug groups exist with no similarities, separate global processes can be used to enable separate within-group information sharing. Additionally, SRS data often include a collapsed, heterogeneous comparator category $J$ (e.g., ``Other drugs''), serving as a baseline group with few, if any, distinct AE signals. In such cases, shrinking all $\lambda_{iJ}$ toward $1$ using a prior such as $\text{Gamma}(\text{shape}=\tau, \text{rate}=\tau)$ with an appropriate hyperparameter $\tau$ can improve signal detection and inference (see Supplement~\ref{ssec:hierarchical-bayes}).

As noted in Section~\ref{sec:DP revisit}, while the DP-Poisson framework is expected to provide some robustness against zero-inflation, a separate ZIP likelihood $n_{ij} \sim \text{ZIP}(\lambda_{ij}E_{ij}, \omega_j)$ may be used if substantial zero inflation is present in the data, or direct inference on structural zero positions/probabilities is required. Complete hierarchical Bayesian representations of our Poisson and ZIP-based models are provided in Supplement~\ref{ssec:hierarchical-bayes}.

\begin{remark}
As a special case of our local-global model, the ``local-only'' DP model of Hu et al.{\cite{hu2015signal}} can be recovered by setting the common local proportion  $\pi=1$, which leads to $z_{ij} = 1$ and hence $\lambda_{ij} = \lambda_{ij}^l$ for all $i = 1, \dots, I$ and $j = 1, \dots, J-1$. Conversely, a ``global-only'' model can be obtained by setting $\pi = 0$, leading to $\lambda_{ij} = \lambda_i^g$ for all $i$ and all $j < J$. We note, however, that this model is considerably more restrictive, as it does not distinguish among $\{\lambda_{i1}, \dots, \lambda_{i,J-1}\}$ for any $i$. This yields the simplified likelihood $n_{ij} \sim \text{Poisson}(\lambda_i E_{ij})$, which produces the sufficient statistic $\sum_{j=1}^{J-1} n_{ij} \sim \text{Poisson}(\lambda_i \sum_{j=1}^{J-1} E_{ij})$ for $\lambda_i$. The model reduction also impacts signal detection. Under both the general local-global and local-only frameworks, hypothesis testing is conducted for each individual $\lambda_{ij}$. In contrast, under the global-only model, inference reduces to testing whether $\lambda_i^g > 1$. Testing $\lambda_i^g > 1$ amounts to assessing whether the aggregated count $\sum_{j=1}^{J-1} n_{ij}$ is large relative to the expected total count $\sum_{j=1}^{J-1} E_{ij}$ under the reduced model. This represents a type of ``global test'' that collapses all drugs into one homogeneous group, whereas the local-global (and local-only) models allow hypothesis testing for each individual drug-AE pair. Consequently, direct comparison between the global-only and local-global models is usually not practically meaningful, as they target different inferential objectives.
\end{remark}

\begin{remark} \label{remark:hierarchical-dp}
Extensive literature exists on local-global DP models for sharing information across groups while preserving DP-based clustering within each group. A common approach employs hierarchical DPs \cite{teh2004sharing}, where local DPs share a global DP as a base distribution. In contrast, our approach leverages a two-component DP mixture framework \cite{antoniak1974mixtures}, separately modeling local and global AE signal strengths and thus offering a more transparent bi-level clustering over AEs and drugs, enabling more flexible estimation of the local and global parameters for greater flexibility in SRS applications.
\end{remark}

\begin{remark} \label{remark:row-vs-col-model}
    In this paper, we focus on drug-based pharmacovigilance, identifying AE (row) signals given one or more drugs (columns in SRS contingency tables). An alternative AE-based approach can be performed to discover drugs strongly associated with a given set of AEs \cite{ding2020evaluation, liu2024pattern}. Our proposed approach can be readily adapted for such analyses: a row-based version of our column-based prior \eqref{eq:DP-local-global}, obtained by switching the roles of $i$ and $j$,  can be used for this purpose.    
\end{remark}

\subsection{Full Bayesian implementation using MCMC} \label{sec:full-bayes-infer}

For full Bayesian inference on our models, we place independent weakly informative proper priors on all model hyperparameters. Along with the exponential prior for $\alpha_j$ and $\alpha$, and half-Cauchy prior for $\beta_j$ and $\beta$ as suggested in Section~\ref{sec:DP revisit}, we consider a similar Cauchy$^{+}$(0, $\psi_\tau$) prior for $\tau$ with $\psi_\tau = 1/2$, and independent Uniform$(0, 1)$ priors for the probability weight parameters, viz., $\pi$ and $\omega_j$ (if included in the model). 

We develop efficient MCMC algorithms for rigorous posterior sampling, approximating the DPs using finite mixtures with $K = O(\log (IJ))$ components \cite{ishwaran2001gibbs}. Each MCMC iteration involves data augmentation to generate latent DP mixture allocations, followed by Gibbs sampling updates for the stick-breaking weights and atoms and the concentration hyperparameters for each local and global DP. The local distribution indicators $z_{ij}$, local distribution proportion $\pi$, AE signal strengths $\lambda_{iJ}$ for the reference drug $J$, and zero-inflation indicators $y_{ij}$ and probabilities $\omega_j$ (if included in the model) also admit standard Gibbs updates. However, the full conditionals for the DP base Gamma hyperparameters ($\beta_j$ and $\beta$) and the reference category shrinkage hyperparameter $\tau$ lack standard forms; for these we employ univariate stepping-out slice sampling \cite{neal2003slice} steps. Detailed MCMC steps are provided in Supplementary Algorithms~\ref{salgo:DPSB}, \ref{salgo:poisson} and \ref{salgo:zip}. For signal detection, we use an FDR-controlled, FNR-optimized hypothesis testing framework based on posterior draws for $\lambda_{ij}$, described next.

\subsection{Signal detection using hypothesis testing} \label{sec:signal-detect}
For signal detection we consider testing the hypotheses: 
\[
  H_{0,ij}: \lambda_{ij} \leq 1 \text{ vs. } H_{1, ij}: \lambda_{ij} > 1 
\] 
for all AEs $i = 1, \dots, I$ and all drugs $j = 1, \dots, J$. Here $\lambda_{ij}$ parametrizes the ratio $n_{ij}/E_{ij}$, and hence the alternative hypothesis $H_{1, ij}$ corresponds a drug-AE pair $(i, j)$ that is deemed as a signal (See Section~\ref{sec:notation-prob-models}). The null hypothesis  $H_{0, ij}$ is rejected in favor of $H_{1, ij}$, identifying the drug-$j$ and AE-$i$ pair to be a signal, if $T_{ij} > k$, where $T_{ij}$ is a test statistic derived from the posterior distribution and $k$ is an appropriate threshold. A common choice for $T_{ij}$ is the marginal posterior quantile $T_{ij} (p)$, defined by $\Pr{\!}_{\text{posterior}}(\lambda_{ij} \leq T_{ij} (p)) = p$ for a small pre-specified quantile level $p \in (0, 1)$ and some $k > 1$, where $\Pr{\!}_{\text{posterior}}(\cdot)$ denotes posterior probability given the data, and can be computed via MCMC draws. This test reflects the intuition that to reject $H_{0, ij}$, \textit{a major bulk} of the posterior distribution of $\lambda_{ij}$ should lie sufficiently above $1$; here the quantile level $p$ quantifies the posterior mass constituting the ``major bulk'' while $k$ measures the degrees of separation from $1$ required to ensure practical significance.  Hu et al.\cite{hu2015signal} suggest using the $5$-th posterior quantiles $T_{ij}(0.05)$ as the test statistic with a rejection threshold of $k = 2$; however, such deterministic thresholds can be overly conservative (curtailing sensitivity) or overly liberal (failing to control false discoveries). Instead, we suggest considering a family of tests $\{T_{ij} (p) > k\}$ with varying $p$ (reasonably small) and $k > 1$, and then selecting the optimal $(\hat p, \hat k)$ that controls the false discovery rate (FDR) while minimizing the false negative rate (FNR). Given $p$ and $k$, FDR and FNR are estimated using the posterior distribution; \cite{ahmed2010false,muller2006fdr} in the expressions below $1(A)$ denotes the indicator of a set $A$,  $\hat{=}$ denotes ``estimated by'', and $T^{\text{obs}}$ is the computed value of $T$):
\begin{equation}
    \label{eq:FDR_FNR}
    \text{FDR}(p, k) \ \hat{=} \ \frac{\sum_{i, j} q_{ij} \ I(T_{ij}(p)^{\text{obs}} > k)}{\sum_{ij} 1(T_{ij}(p)^{\text{obs}} > k)}, \quad
\text{FNR}(p, k) \ \hat{=} \ \frac{\sum_{i, j} (1 - q_{ij})\ 1(T_{ij}(p)^{\text{obs}} \leq  k)}{\sum_{ij} 1(T_{ij}(p)^{\text{obs}} \leq k)},
\end{equation}
\noindent where $q_{ij} = \Pr{\!}_{\text{posterior}}(\lambda_{ij} \leq 1)$ is the posterior probability of $H_{0,ij}$, computable from MCMC draws. A grid search is used to determine the optimal $(\hat p, \hat k)$ that ensures $\text{FDR}(\hat p, \hat k) \leq \alpha$ for a pre-specified level $\alpha$ while minimizing $\text{FNR}$. As a default choice, we consider a $k$-grid ranging over $[1.1, 3]$ and a $p$-grid over $[0.01, 0.10]$ (see Supplement{~\ref{ssec:pFDR_calculation}} for details on the grid construction). A more adaptive grid, particularly for $k$, may be constructed by monitoring the maximum of the posterior draws of $\lambda_{ij}$ across all $i$ and $j$; however, our simulation-based sensitivity analyses (see Remark{~\ref{remark:sensitivity-analysis})} suggest that the default grid is generally adequate for signal detection.

In practice, drug-AE pairs with $n_{ij} = 1$ are often deemed non-signals, even if $\lambda_{ij} \gg 1$ with high posterior probability, leading to $T_{ij}(p) > k$ for suitably chosen $p$ and $k$. To prevent their detection as signals, the rejection region can be modified to $\{T_{ij}(p) > k \text{ and } n_{ij} > 1\}$.

\section{SIMULATION ASSESSMENTS} \label{sec:simulation}

In this section, we conduct extensive simulations to assess the signal detection performance of our local-global DP Poisson and ZIP models under a wide array of data-generating settings. We follow the SRS contingency table generation approach previously used in the literature \cite{chakraborty2022use, liu2024pattern}. Specifically, to generate random SRS contingency tables of dimension $I \times J$, we first define a prespecified \textit{true} signal strength matrix (described below) $\Lambda^0 = ((\lambda_{ij}^0))$ encoding signals, structural zeros, non-signals, and between-drug dependency patterns. Using a reference $I \times J$ contingency table (described next) from a real-world SRS database, we extract row and column totals $\{\boldsymbol{n}_{i\bullet}: i=1, \dots, I\}$ and $\{\boldsymbol{n}_{\bullet j}: j = 1, \dots, J\}$, respectively, to facilitate random contingency table generation. Next, we generate $\boldsymbol{p}_{i\bullet} \sim \text{Dirichlet}(\boldsymbol{n}_{i\bullet})$ and $\boldsymbol{p}_{\bullet j} \sim \text{Dirichlet}(\boldsymbol{n}_{\bullet j})$ for all $i$, $j$ defining the random probability matrix $\boldsymbol{p} = ((p_{ij}))$ with $p_{ij} \propto \lambda_{ij}^0 \ \boldsymbol{p}_{i\bullet} \ \boldsymbol{p}_{\bullet j}$ and $\sum_{i, j} p_{ij} = 1$. Finally, we generate a random SRS contingency table by sampling from $\text{Multinomial}(\boldsymbol{n}_{\bullet \bullet}; \boldsymbol{p})$, collapsing any generated $n_{ij} = 1$ corresponding to a true signal (as defined by $\lambda^0_{ij}$) to $n_{ij} = 0$ afterward to eliminate signal detection at $n_{ij} = 1$.

\paragraph*{Reference SRS dataset to compute row and column totals} 
We use the FDA FAERS database (Q3 2014-Q4 2020), which catalogs a total of $63{,}976{,}610$ reported cases \cite{chakraborty2023likelihood}, to construct our reference SRS contingency table. We focus on six statin medications: Atorvastatin, Fluvastatin, Lovastatin, Pravastatin, Rosuvastatin, and Simvastatin, collapsing all other drugs into a comparator ``Other drugs'' category. For expository purposes, and to create a sufficiently large SRS contingency table, we focused on $I=1491$ preferred terms (PTs) as AEs that are common (appear at least once) for these statin drugs as well for a separate group of drugs unrelated to statin,  namely, Gadolinium-based contrast agents (GBCA) drugs \cite{chakraborty2022use}. All other PTs recorded in the database within this period were collapsed into a heterogeneous category labeled ``Other AE'', yielding a $1492 \times 7$ contingency table summarizing the report counts for each drug-AE pair. Alongside this larger setting, we also consider a smaller setting involving a subset of this dataset with $I=46$ relevant AEs as considered in prior work{\cite{chakraborty2022use, huang2011likelihood}} and a collapsed Other AE cateogory;  see Supplement~{\ref{ssec:statin1491}}.

\paragraph*{Constructing the true signal strength matrix $\Lambda^0$} To generate $\Lambda^0$, we define two signal patterns: (1) \textit{fixed row signals}, where all drugs (columns) exhibit signals for the same AEs (PTs; rows), and (2) \textit{random signals}, where signals appear on randomly chosen AEs (different row for each drug). A larger number of fixed row signals increase the between-drug associations, measured via correlations between $\lambda_{ij}^0$ and $\lambda_{ij'}^0$ for different statin drugs $j$ and $j'$, while more random signals weaken them. The matrix $\Lambda^0$ is constructed by specifying fixed row signals and embedding additional random signals in the remaining rows across all statin columns. Signal strengths in these cells range from $1 < \lambda_{ij}^0 \leq 3$ (see Table~\ref{tb:simulation_scenarios}). The remaining cells $(i', j')$ are designated as non-signals with $\lambda_{i'j'}^0 = 1$. To introduce zero inflation, we prespecify \textit{true} $\omega_j^0$ ($0$ or $0.30$; Table~\ref{tb:simulation_scenarios}), and set $\omega_j^0 I$ randomly selected non-signal cells to $\lambda_{ij}^0 = 0$. This approach yields a controlled $\Lambda^0$ to generate random SRS contingency tables. Within this setup, we consider four simulation scenarios with varied range of between-column correlations in $\Lambda^0$, each with and without zero-inflation (see Table~\ref{tb:simulation_scenarios}).

\begin{table}[ht!]
\centering

\caption{Simulation scenarios describing the construction of the true signal strength matrix $\Lambda^0$ encoding signals, non-signals, zero-inflation, and between-drug associations. The first two rows in the table show the number of fixed-AE/row signals and the number of random signals per (statin) column in $\Lambda^0$. The zero inflation rate (third row in the table) shows the drug-specific true zero-inflation probability $\omega_j^0$ in each simulation scenario. The comparator ``Other drugs'' column has all non-signal $(\lambda_{ij}^0 = 1)$ cells. Finally, the fourth row in the table shows the average Kendall's $\tau$ between $\lambda_{ij}^0$ and $\lambda_{ij'}^0$ across all statin drug pairs $j$ and $j'$ (excluding the ``Other drugs'' category) as a measure of pairwise association of their AE signal strengths.}

\label{tb:simulation_scenarios}
\begin{threeparttable}
\begin{tabular}{lcccccccc}
\hline
Simulation Scenario                      & 0a   & 0b   & 1a   & 1b   & 2a   & 2b   & 3a   & 3b   \\ \hline
No. of fixed row signals                 & 1    & 1    & 3    & 3    & 10   & 10   & 20   & 20   \\
No. of random signals per column         & 30   & 30   & 20   & 20   & 10   & 10   & 3    & 3    \\
Zero-inflation rate per column           & 0    & 0.3  & 0    & 0.3  & 0    & 0.3  & 0    & 0.3  \\
Average Kendall's $\tau$ between column pairs& $\leq$0.01 & $\leq$0.01 & 0.12 & 0.12 & 0.49 & 0.49 & 0.77 & 0.77 \\ \hline
\end{tabular}
\begin{tablenotes}
\item A common signal strength value $\lambda^{0, \text{signal}}$ is assigned to all signal cells in a given simulation setting, and we vary $\lambda^{0, \text{signal}} \in \{1.01, 1.1, 1.2, 1.3, 1.4, 1.5, 1.7, 1.9, 2, 2.5, 3\}$.
\end{tablenotes}
\end{threeparttable}
\end{table}

\paragraph*{Replicated Data Generation, Model Fitting, and signal detection}  For each constructed signal strength matrix $\Lambda$ (Table~\ref{tb:simulation_scenarios}), and using the random contingency table generation mechanism described earlier, we generate {$R = 1{,}000$} independent random contingency table replicates. In each replicate, we fit our models via MCMC sampling (Supplementary Algorithms~\ref{salgo:DPSB}, \ref{salgo:poisson} and \ref{salgo:zip}) with $10{,}000$ post-burn-in draws after discarding the first $5,000$ burn-in draws; see Supplement~\ref{ssec:computational_details} for additional computational details, including starting points for the MCMC algorithms. From the computed posteriors, we conduct signal detection using hypothesis testing (Section~\ref{sec:signal-detect}).
 
We compare our signal detection performance against existing Bayesian SRS data mining methods, viz., BCPNN, GPS, and the \textit{local-only} DP model of Hu et al.\cite{hu2015signal} (henceforth DP Hu et al).  For our models and DP Hu et al., signal detection uses an optimized posterior quantile level $\hat{p}$ (fixed at 0.05 for DP Hu et al.) and threshold $\hat{k}$ ensuring FDR $\leq 0.05$ while minimizing FNR (Section~\ref{sec:signal-detect}), with the additional constraint $n_{ij} > 1$ for our methods. To ensure fair FDR control ($\leq 0.05$) in BCPNN and GPS, we leverage the independence of signal strengths in the (empirical) Bayes posteriors for these two models and compute Benjamini-Hochberg-adjusted \cite{muller2006fdr} posterior probabilities of being a non-signal; a drug-$j$ and AE-$i$ pair is then classified as a signal if the adjusted probability is $\leq 0.05$. Additionally, we evaluate the state-of-the-art frequentist ZIP-based extended (pseudo) LRT \cite{chakraborty2022use, huang2011likelihood, ding2020evaluation}, which ensures FDR control ($\leq 0.05$) with high sensitivity while addressing zero inflation in the counts when present.

\begin{figure}[htpb]
\centering
\includegraphics[width=\linewidth]{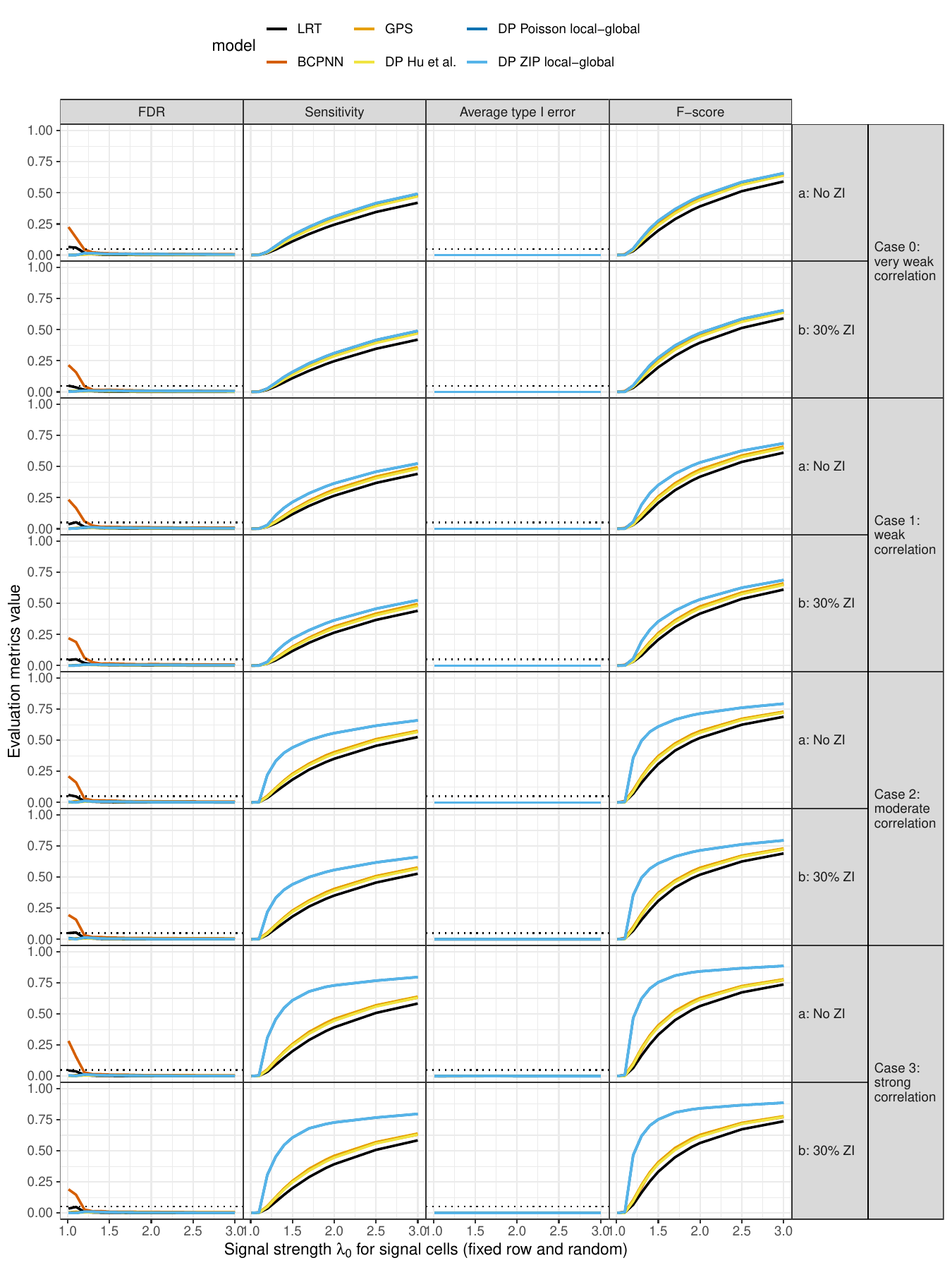} 
\caption{Simulation results: the horizontal axis shows the true signal strength $\lambda_0$ for signal cells, the vertical axis shows the value of evaluation metrics. Each column presents one metric and each row presents one simulation scenario in Table~\ref{tb:simulation_scenarios}.}
\label{fg:simulation}
\end{figure}

We assess signal detection for all methods using replication-based FDR, sensitivity, average Type I error, and F-score (Figure~\ref{fg:simulation}). For each method, these metrics are computed as FDR = $R^{-1} \sum_r \bigl(\text{FP}_r / (\text{FP}_r + \text{TP}_r)\bigr)$, sensitivity = $R^{-1} \sum_r \bigl(\text{TP}_r / (\text{TP}_r + \text{FN}_r)\bigr)$, average type I error = $R^{-1} \sum_r \bigl(\text{FP}_r / \text{TN}_r\bigr)$, and F-score = $R^{-1} \sum_r \bigl(2\text{TP}_r / (2\text{TP}_r + \text{FN}_r + \text{FP}_r)\bigr)$ where $\text{TP}_r$ and $\text{FN}_r$ denote the number of \textit{true signals} that are discovered and not discovered respectively, and $\text{FP}_r$ and $\text{TN}_r$ are the number of \textit{true non-signals} discovered and not discovered in replicate $r$ based on the signal strength values as cataloged in the corresponding \textit{true} $\Lambda$.

Our results show that across all settings, the FDR and average type I error are strictly controlled by all methods except BCPNN, which appears to entirely lose control of FDR under the near-null setting of $\lambda^{0,\text{signal}} = 1.01$, despite the use of a Benjamini-Hochberg-based FDR adjustment during signal detection.

Focusing on sensitivity, the first four rows of Figure~\ref{fg:simulation} show comparable (when $\lambda^{0, \text{signal}} < 1.5$) or slightly better (when $\lambda^{0, \text{signal}} \geq 1.5$) sensitivity for our local-global methods compared to all other methods when the there is very weak  ($\tau \leq 0.01$; Cases 0a, b) or  weak ($\tau = 0.12$; Cases 1a,b) between-drug associations. Under moderate between-drug association ($\tau = 0.49$; Cases 2a,b), our methods perform noticeably better in detecting more true signals, achieving up to twice the sensitivity, particularly for weak or moderate true signal strengths ($1.1 \leq \lambda^{0, \text{signal}} \leq 1.7$). The gain is even further pronounced under high between-drug association ($\tau = 0.77$; Cases 3a,b), where our methods attain sensitivities of $\geq 0.70$ with only moderate true signal strength ($\lambda^{0, \text{signal}} \geq 1.7$), while all competitors fail to reach $0.70$ sensitivity even for $\lambda^{0, \text{signal}} = 3$.  These results highlight the substantial boost in signal detection from effectively leveraging between-drug similarities as facilitated by our models.

The Poisson and ZIP-based DP local-global models perform similarly across all scenarios regardless of the presence of zero-inflated observed counts (sub-cases a and b), due to the flexibility of the DP-based nonparametric framework (see Section~\ref{sec:DP revisit}).

The results from the smaller setting involving $I=46$ AEs are essentially similar to those obtained in the larger setting; see Supplementary Figure~{\ref{sfg:simulation_statin46}} in Supplement~{\ref{ssec:simu_statin46}} for further details. All computations were performed on the cluster computing facility at the Center for Computational Research at the University at Buffalo, with each process running on 16 GB RAM and a clock speed of 1.6-3 GHz. The median computation time per replication for our DP local-global methods (both Poisson and ZIP) is approximately 48 minutes for the larger setting with 1491 AEs, compared with 0.66, 0.02, 0.42, and 26 minutes for BCPNN, GPS, LRT, and DP Hu et al., respectively. For the smaller 46-AE dataset, our methods require about 2 minutes per replication, versus 0.02, 0.002, 0.07, and 0.7 minutes for the same competing approaches. The higher computational cost reflects our substantially richer modeling assumptions: except for DP Hu et al., the competing methods achieve highly efficient computation through partial analytical tractability but do not provide flexible, regularized estimation of signal strength parameters with full uncertainty quantification, nor do they capture complex between-drug associations in AE signal patterns. Relative to DP Hu et al., our methods introduce only one additional global DP component and the local-global mixture parameters, incurring modest extra cost while delivering substantial gains in sensitivity and signal detection, as demonstrated in our simulations.

\begin{remark} \label{remark:sensitivity-analysis}
Leveraging the simulation setup described as above, we performed additional sensitivity analyses to examine the robustness of the our methods with respect to the choice of hyperparameters, including the prior hyperparameters $\psi_\alpha$ and $\psi_\beta$, as well as the grid specification for the hypothesis testing threshold $k$. For each of $\psi_\alpha$ and $\psi_\beta$, we considered two alternative choices representing values toward the lower and upper extremes of the weakly informative range along with the default $\psi_\alpha = 3$ and $\psi_\beta = 1/2$. Specifically, for $\psi_\alpha$ we considered $\psi_\alpha = 1$ (placing $\approx 64\%$ mass in $(0,1)$) and $\psi_\alpha = 5$ ($\approx 99\%$ mass in $(0,1)$), and for $\psi_\beta$ we considered $\psi_\beta = 1/7$ ($\approx 90\%$ mass in $(0, 1)$) and $\psi_\beta = 6$ (only $10\%$ mass in $(0, 1)$). For the $k$-grid construction, we considered a more adaptive approach monitoring the computed maximum of all $\lambda_{ij}$ posterior draws instead of the default grid on $[1.1, 3]$. Across all settings, the resulting performance curves were virtually identical to those obtained under the default specifications, indicating that the proposed method is largely stable across a range of reasonable prior choices and grid settings. A detailed description of the sensitivity analysis results is provided in the Supplement~{\ref{ssec:simulation_krange}}.
\end{remark}

\section{REAL DATA ANALYSIS} \label{sec:data analysis}

\begin{figure}[h]
\centering
\includegraphics[height=0.85\textheight]{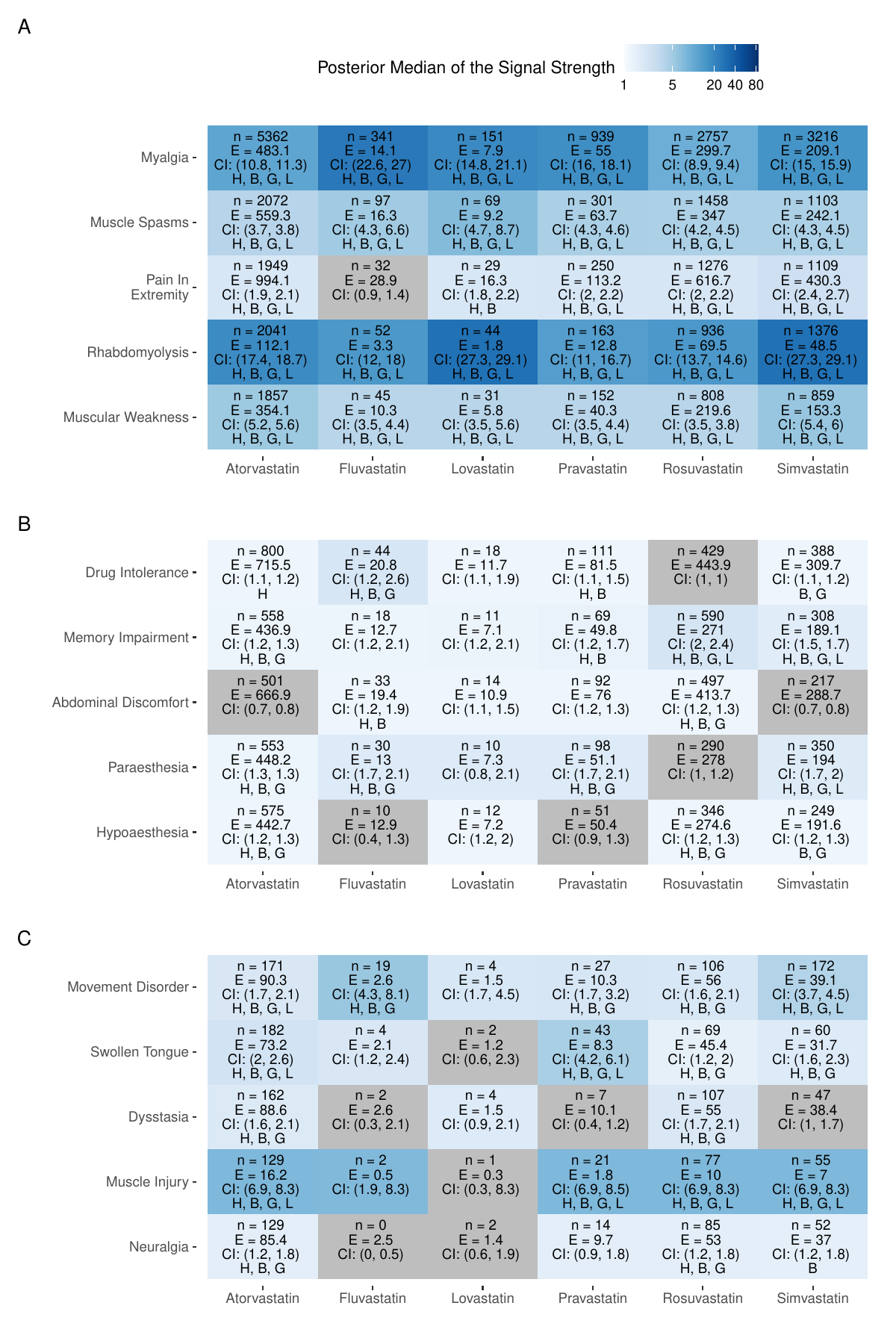} 
\caption{Heatmaps of 15 selected PTs. Cells not detected as signals by our method are colored gray. White-to-blue shades represent the median of the posterior of signal strength parameter \(\lambda_{ij}\), with darker (lighter) blue indicating a stronger (weaker) signal strength. Panel A contains 5 common PTs among all statin drugs. Panel B contains 5 semi-rare PTs with new discovered signals by our proposed method. Panel C contains 5 rare PTs with new discovered signals by our proposed method. Count $n$, expected count $E$, and the 90\% credible interval of the posterior distribution of $\lambda_{ij}$ are shown in each cell. The letters ``H'', ``B'', ``G'', and ``L'' indicate signal detection by the DP Hu et al., BCPNN, GPS, and likelihood ratio test, respectively.}
\label{fg:statin_heatmap}
\end{figure}

We use the statin data from Section~\ref{sec:simulation} to demonstrate our method's utility in real-world SRS data mining. We fit our DP-Poisson and DP-ZIP local-global models to the dataset, setting the finite mixture approximation of the DP with $K= 20$ components,
and using the same MCMC and test statistic generation settings as used for simulated data analysis in Section~\ref{sec:simulation}). We compare the detected signals against those from the DP Hu et al., BCPNN, GPS, and LRT. Table~\ref{tb:data_analysis} presents the number of signals identified across statin drugs by all these methods. The results indicate very similar performance for our Poisson and ZIP local-global models, both of which detect a substantially larger number of signals than all competing methods.

\begin{table}[]
\caption{Summary table of discoveries from our proposed DP local-global models, and comparing them to the other existing methods. The count in each cell represents the number of AE signals detected by the method (along the columns), first separately for each drug and then in total (rows). The numbers in parentheses represent the differences in the number of discoveries made by each method when compared to our DP Poisson local-global method.}
\label{tb:data_analysis}
\begin{tabular}{lrrrrrr}
\hline
Drug         & DP Poisson local-global & DP ZIP  local-global    & DP Hu et al. & BCPNN     & GPS       & LRT        \\ \hline
Atorvastatin & 308                        & 306 (2)                    & 307 (1)                          & 234 (74)                  & 203 (105)               & 126 (182)               \\
Fluvastatin  & 136                        & 136 (0)                    & 122 (14)                         & 68 (68)                   & 60 (76)                 & 26 (110)                \\
Lovastatin   & 90                         & 94 (-4)                    & 59 (31)                          & 37 (53)                   & 30 (60)                 & 11 (79)                 \\
Pravastatin  & 226                        & 227 (-1)                   & 210 (16)                         & 122 (104)                 & 91 (135)                & 35 (191)                \\
Rosuvastatin & 314                        & 314 (0)                    & 297 (17)                         & 228 (86)                  & 195 (119)               & 110 (204)               \\
Simvastatin  & 314                        & 317 (-3)                   & 257 (57)                         & 220 (94)                  & 183 (131)               & 94 (220)                \\
Total        & 1388                       & 1394 (-6)                  & 1252 (136)                       & 909 (479)                 & 762 (626)               & 402 (986)               \\ \hline
\end{tabular}
\end{table}

To further examine the signal detection results, we focus on our DP Poisson local-global model and zoom in on 15 selected PTs, whose signal strength estimates are plotted using heatmaps in Figure~\ref{fg:statin_heatmap}. The figure is divided into three panels based on the frequencies of the AEs, and the colors in the heatmaps indicate either non-detection (gray) or the estimated signal strengths measured through the posterior medians of $\lambda_{ij}$ (white-to-blue gradient with a darker blue signifying a stronger signal strength). Panel A displays the five most common AEs with the largest frequencies (total counts for the AEs across all six statin drugs being greater than $2000$) and the strongest estimated signal strengths (having the largest posterior medians of $\lambda_{ij}$ across all statin drugs $j$) that are detected as a signal for at least one statin drug by our method. The AEs displayed in this panel are all well-documented muscle-related conditions, such as rhabdomyolysis \cite{antons2006clinical, tournadre2020statins} and myopathy \cite{ballantyne2003risk, abd2011statin}, and their detection by our method is consistent with the performance of all competing methods. Panel B contains five semi-rare AEs (with observed counts across statin drugs ranging between 500 and 2000) that exhibit the strongest estimated signal strengths. This category primarily includes some specific types of disorders, such as nervous system disorders, and interestingly, several of these detected signals are not identified by any competing method for the newer or less frequently used statin drugs. For example, unlike other methods, our model detects lovastatin-induced drug intolerance as a signal, which has been flagged as a potential side effect of lovastatin\cite{ahmad2014statin}. Additionally, our method detects signals for paraesthesia and hypoaesthesia, symptoms of peripheral neuropathy, associated with lovastatin use; this finding is also consistent with previous research \cite{chong2004statin, wang2021association}. By appropriately borrowing strengths from the more commonly used statins, our method is successfully able to identify potential lovastatin-induced AE signals in these specific examples.  Finally, Panel C focuses on five rare AEs (with observed counts across statin drugs being lower than 500) with the strongest estimated signal strengths. Within this group, our method identifies several AEs related to neurological and movement issues, such as movement disorder and swollen tongue, and some other mixed disorders. Many of these AEs were not originally discovered in clinical trials, but later were flagged as potential signals in specific, targeted pharmacovigilance investigations. For example, fluvastatin-induced swollen tongue, also known as angioedema, has been recently recorded \cite{kumar2024fluvastatin}. Similar to the semi-rare AEs displayed in Panel B, our method is able to identify these rare AE signals for the less frequently used drug fluvastatin by appropriately sharing AE signal strength information across all statin drugs.

\section{DISCUSSION}

We propose nonparametric Bayesian models and their MCMC implementation for SRS data mining, balancing flexibility with shrinkage while capturing between-drug associations. Signal detection is performed through hypothesis tests that control FDR and minimize FNR, both estimated from MCMC posteriors. Our approach achieves significantly higher sensitivity than existing methods, particularly for rare AEs and/or newer drugs, with small observed counts. Application to the statin dataset demonstrates our model's superior performance, which detects more plausible signals involving rare AEs and/or newer drugs than competing methods while reliably identifying well-established AEs. By enabling between-drug information sharing, our model can also discover potential signals among rare AEs and provide more nuanced detection for semi-rare AEs. 

There are several limitations with our current approach that motivate interesting future directions. First, we focus here primarily on signal detection, i.e., to identify critical AEs of concern for each drug. The clusters of AEs produced by our DP priors have not been investigated but may reveal additional, deeper insights by identifying groups of ``similar AEs''. A reformulation via mixed membership models could further enhance interpretability by allowing the same AE to appear in multiple clusters/groups. Second, while we consider data-driven characterization of similarities in AE signal strengths between drugs, a more flexible approach acknowledging known Drug-Drug Interactions (DDI \cite{poleksic2019database}), possibly via appropriate prior distributions, may further enhance signal detection. Finally, our current framework aggregates drug-AE data across time points for a cross-sectional analysis. Extending the model to accommodate temporal variations in SRS data could enable formal identification of evolving AE patterns.

\noindent \textbf{Acknowledgment.} The authors are deeply grateful to Prof. Marianthi Markatou at the University at Buffalo for her insightful comments and helpful suggestions on an earlier version of the manuscript.  

\bibliographystyle{plain}
\bibliography{main}

\clearpage
\appendix
\begin{center}
  {\Large \bfseries Supplementary Material}
\end{center}
\renewcommand{\thesection}{S.\arabic{section}}
\renewcommand{\thefigure}{S.\arabic{figure}}
\renewcommand{\thetable}{S.\arabic{table}}
\renewcommand{\thealgorithm}{S.\arabic{algorithm}}
\setcounter{section}{0}
\setcounter{figure}{0}
\setcounter{table}{0}
\setcounter{algorithm}{0}

\noindent This supplement provides additional details about the DP local-global models, detailed steps for the MCMC algorithms implementing our models, additional computation details, and additional simulation results. Sections, tables, and figures in this supplement are labeled with a prefix ``S.'', such as, Section S.1, Table S.1, and Figure S.1, etc.

\section{ADDITIONAL DETAILS FOR THE DP LOCAL-GLOBAL FRAMEWORK}

In this section, we generalize the model to encompass different variants depending on the direction of information sharing. Moreover, we provide additional details of the DP local-global framework. And we propose a simpler shrinkage model as an alternative for the ``Other drugs'' as described in Section~\ref{sec:DP new}.

\subsection{Complete hierarchical Bayesian representation of our models}
\label{ssec:hierarchical-bayes}

In Section~\ref{sec:DP new} in the main text, we present the concept of the DP Poisson local-global model. The fully formulated hierarchical representations are provided in the following two sections for the DP Poisson and the DP ZIP local-global model.

\vspace{-0.1in}
\subsubsection{DP Poison local-global model}

A hierarchical representation of the complete model, including vague priors for the hyperparameters, is presented as follows. 

\begin{gather*}
  \text{likelihood:} \\
   n_{ij} \sim \text{Poisson}(\lambda_{ij} E_{ij}) \\  \\
   \text{DP local-global prior for a non-baseline drug } j \in \{1, \dots, J-1\}\!: \\ 
   \lambda_{ij} \sim z_{ij} \lambda_{ij}^l + (1 - z_{ij})  \lambda_i^g \\
   \lambda_{ij}^l \sim D_j; \ \lambda_i^g \sim D \\
   z_{ij} \sim \text{Bernoulli}(\pi) \\  \\
   \text{DP local prior for drug-$j$:} \\
   D_j \equiv \sum_{h=1}^{\infty} \eta_{hj} \delta_{\{\theta_{hj}\}} \approx \sum_{h=1}^{K} \eta_{hj} \delta_{\{\theta_{hj}\}} \sim \text{DP}(\alpha_j, G(\beta_j)) \\
   \theta_{hj} \sim G(\beta_j) \equiv \text{Gamma}(\text{shape} = \beta_j, \text{rate} = \beta_j) \\
   \eta_{1j} = v_{1j}, \ \eta_{hj} = v_{hj} \prod_{t=1}^{h-1}(1 - v_{tj}), h \geq 2  \\
   v_{hj} \sim \text{Beta}(1, \alpha_j), h = 1, \dots, K-1; \ v_{Kj} = 1  \\  \\
   \text{DP global prior:} \\
   D \equiv \sum_{h=1}^{\infty} \eta_{h} \delta_{\{\theta_{h}\}} \approx \sum_{h=1}^{K} \eta_{h} \delta_{\{\theta_{h}\}}  \sim DP(\alpha, G(\beta)) \\
   \theta_{h} \sim G(\beta) \equiv \text{Gamma}(\text{shape} =\beta, \text{rate} = \beta) \\
   \eta_{1} = v_{1}, \ \eta_{h} = v_{h} \prod_{t=1}^{h-1}(1 - v_{t}), h \geq 2 \\
   v_{h} \sim \text{Beta}(1, \alpha), h = 1, \dots, K-1; \ v_{K} = 1\\ \\
   \text{Gamma shrinkage prior for a baseline drug-$J$:} \\
   \lambda_{iJ} \sim \text{Gamma}(\text{shape} = \tau, \text{rate} = \tau) \\ \\
   \text{vague proper priors for hyperparameters:} \\
   \pi^g \sim \text{Uniform}(0, 1) \\
   \alpha_j, \alpha \sim \text{Gamma}(\text{shape} = 1, \text{rate} = 3)\\
   \beta_j, \beta \sim \text{Cauchy}^{+}(0, 0.5) \\
   \tau \sim \text{Cauchy}^{+}(0, 0.5)
\end{gather*}

\subsubsection{DP ZIP local-global model}

A hierarchical representation of the complete model, including vague priors for the hyperparameters, is presented as follows. 

\begin{gather*}
  \text{likelihood:} \\
   n_{ij} \sim y\text{Poisson}(\lambda_{ij} E_{ij}) + (1 - y) \delta_{\{0\}} \\  
   y \sim \text{Bernoulli}(\omega_j) \\ \\
   \text{DP local-global prior for a non-baseline drug } j \in \{1, \dots, J-1\}\!: \\ 
   \lambda_{ij} \sim z_{ij} \lambda_{ij}^l + (1 - z_{ij})  \lambda_i^g \\
   \lambda_{ij}^l \sim D_j; \ \lambda_i^g \sim D \\
   z_{ij} \sim \text{Bernoulli}(\pi) \\  \\   
   \text{DP local prior for drug-$j$:} \\
   D_j \equiv \sum_{h=1}^{\infty} \eta_{hj} \delta_{\{\theta_{hj}\}} \approx \sum_{h=1}^{K} \eta_{hj} \delta_{\{\theta_{hj}\}} \sim \text{DP}(\alpha_j, G(\beta_j)) \\
   \theta_{hj} \sim G(\beta_j) \equiv \text{Gamma}(\text{shape} = \beta_j, \text{rate} = \beta_j) \\
   \eta_{1j} = v_{1j}, \ \eta_{hj} = v_{hj} \prod_{t=1}^{h-1}(1 - v_{tj}), h \geq 2  \\
   v_{hj} \sim \text{Beta}(1, \alpha_j), h = 1, \dots, K-1; \ v_{Kj} = 1  \\  \\
   \text{DP global prior:} \\
   D \equiv \sum_{h=1}^{\infty} \eta_{h} \delta_{\{\theta_{h}\}} \approx \sum_{h=1}^{K} \eta_{h} \delta_{\{\theta_{h}\}}  \sim DP(\alpha, G(\beta)) \\
   \theta_{h} \sim G(\beta) \equiv \text{Gamma}(\text{shape} =\beta, \text{rate} = \beta) \\
   \eta_{1} = v_{1}, \ \eta_{h} = v_{h} \prod_{t=1}^{h-1}(1 - v_{t}), h \geq 2 \\
   v_{h} \sim \text{Beta}(1, \alpha), h = 1, \dots, K-1; \ v_{K} = 1\\ \\
   \text{Gamma shrinkage prior for a baseline drug-$J$:} \\
   \lambda_{iJ} \sim \text{Gamma}(\text{shape} = \tau, \text{rate} = \tau) \\ \\
   \text{vague proper priors for hyperparameters:} \\
   \pi \sim \text{Uniform}(0, 1) \\
   \alpha_j, \alpha \sim \text{Gamma}(\text{shape} = 1, \text{rate} = 3)\\
   \beta_j, \beta \sim \text{Cauchy}^{+}(0, 0.5) \\
   \tau \sim \text{Cauchy}^{+}(0, 0.5)
\end{gather*}

\section{COMPUTATIONAL DETAILS} \label{ssec:computational_details}
In this section, we first provide MCMC algorithms implementing our models in Section~\ref{ssec:MCMC_algo}. The MCMC algorithms are derived using finite mixture approximation of the underlying DP processes \citep{ishwaran2002exact, fruhwirth2019here}, and combine Gibbs sampling updates with stepping-out slice sampling updates \citep{neal2003slice} for various parameters. Some additional notes on the MCMC implementation, including initial value generation for our MCMC algorithms are then provided in Section~\ref{ssec:MCMC_initial}. Next we provide detailed notes on the grid-search method identifying optimal quantile-level $p$ and threshold $k$ used for our hypothesis tests for signal detection (Section~\ref{ssec:pFDR_calculation}).  The JAGS code \citep{plummer2003jags} for \citet{hu2015signal}'s models is shown as a reference in Section~\ref{ssec:JAGS code}. The example statin dataset is included in Section~\ref{ssec:statin1491}.

In our notation below, $\text{DPSB}_K(\bm \eta = (\eta_1, \dots, \eta_K), \bm \theta = (\theta_1, \dots, \theta_K), \alpha, \beta)$ denotes parameters associated with the model $\sum_{h=1}^K \eta_h 1_{\{\theta_h\}} \sim \text{DP}(\alpha, G(\beta))$, comprising the stick-breaking weights $\bm \eta = (\eta_1, \dots, \eta_K)$ and atoms $\bm \theta = (\theta_1, \dots, \theta_K)$ and DP concentration hyperparamter $\alpha$ and base (gamma) distribution hyperparameter $\beta$.

\subsection{MCMC algorithms}
\label{ssec:MCMC_algo}

\begin{algorithm}[H]

\caption{One Iteration of the MCMC algorithm updating $\text{DPSB}_{K}(\boldsymbol{\eta}, \boldsymbol{\theta}, \alpha, \beta)$ parameters from its posterior given data $(n_{i}, E_{i}, z_{i})$ under the model: $n_{i} \sim \text{Pois}(\lambda_{i} E_{i})$, $\lambda_{i} \sim \text{DPSB}_K(\boldsymbol{\eta}, \boldsymbol{\theta}, \alpha, \beta)$, $\alpha \sim \text{Exponential}(\text{rate}=\psi_\alpha)$, $\beta \sim \pi_\beta(\beta)$, and $z_i$ indicating if observation $i$ contributes to the model ($z_i = 1$) or not ($z_i = 0$); $i = 1, \dots, N$.} 
\label{salgo:DPSB}

\begin{enumerate}
    \item Generate allocations $\bm S = (S_{1}, \dots, S_{N})$ using 
      \[
        P(S_i = h \mid \boldsymbol{\eta}, \boldsymbol{\theta}, \boldsymbol{z}, \boldsymbol{n}, \boldsymbol{E}) = \frac{\eta_{h} \{ f_{\text{Poisson}}(n_i \mid \theta_{h}E_{i})\}^{z_{i}}}{\sum_{h'=1}^{K}\eta_{h'} \{f_{\text{Poisson}}(n_i \mid \theta_{h'}E_{i})\}^{z_{i}}} 
    \]  

    for $h = 1, \dots, K$. Here $f_{\text{Poisson}}(n \mid \theta)$ denotes the probability mass function of the Poisson($\theta$) distribution evaluated at $n$. 
    
    \item Generate $\theta_{1}, \dots, \theta_{K}$ independently from
      \[
        \theta_{h} \mid \boldsymbol{S}, \alpha, \boldsymbol{n}, \boldsymbol{E}, \boldsymbol{z} \sim \text{Gamma}\!\Biggl(\text{shape} = \sum_{i=1}^{N} n_{i} z_{i} 1(S_{i} = h) + \alpha,  \text{rate} = \sum_{i=1}^{N}  E_{i} z_{i} 1(S_i = h) + \alpha \Biggr)
    \]

        \begin{enumerate}[label*=\arabic*.]
            \item Calculate $\lambda_{i} = \theta_{S_{i}}$, $i = 1, \dots, N$
        \end{enumerate}

    \item Generate $\beta$ from the density
        \[
            \pi(\beta \mid \boldsymbol{\theta}) \propto \prod_{h = 1}^{K} f_{\text{Gamma}}(\theta_h \mid \text{shape} = \beta, \text{rate} = \beta) \pi_{\beta}(\beta)
        \]
        Here $f_\text{Gamma}(\theta \mid \text{shape} = \beta, \text{rate} = \beta)$ denotes the probability density function of the Gamma($\beta, \beta$) distribution evaluated at $\theta$. $\pi(\beta)$ is the prior distribution of $\beta$.
        Direct random sampling from this density is not feasible. Instead, we use stepping-out slice sampling \citep{neal2003slice} to generate Markov chain draws for $\beta$.
       
    \item Generate $v_{1}, \dots, v_{K-1}$  independently from:
        \[
            v_{h} \mid \boldsymbol{S}, \alpha \sim \text{Beta} \left( 1 + \sum_{i=1}^{N} 1(S_{i} = k), \alpha + \sum_{i=1}^{N} 1(S_{i} > k) \right).
        \]
        Then set $v_K = 1$ and calculate $\eta_{1}, \dots, \eta_{K}$ from $v_{1}, \dots, v_{K}$ using
            \[
                 \eta_{1} = v_{1}, \quad \eta_h = v_{h} \prod_{t=1}^{h-1} (1 - v_{t}), \quad h = 2, \dots, K.
            \]
        
    \item Generate $\alpha$ from
        \[
            \alpha \mid v_{1}, \dots, v_{K} \sim \text{Gamma} \left( \text{shape} = K, \text{rate} = -\sum_{h=1}^{K-1} \log(1 - v_{h}) + \psi_{\alpha} \right).
        \]
  \end{enumerate}
\end{algorithm}

\begin{algorithm}[H]

\caption{One Step Iteration of DP Poisson local-global column model (Given data $\{(n_{ij}, E_{ij}, z_{ij}): i = 1, \dots, I; j = 1, \dots, J\}$ )}
\label{salgo:poisson}

    \begin{enumerate}
        \item Generate global $\text{DPSB}_{K}(\boldsymbol{\eta}, \boldsymbol{\theta}, \alpha, \beta)$ parameters using Algorithm~\ref{salgo:DPSB}, with $\{(\tilde{n}_{i}, \tilde{E}_{i}, 1(\sum_{j=1}^{J-1} z_{ij} < J-1) ): i = 1, \dots, I\}$ as data where
        \[
        \tilde{n}_{i} = \sum_{j=1}^{J-1} n_{ij}(1 - z_{ij}), \quad \tilde{E}_{i} = \sum_{j=1}^{J-1} E_{ij}(1 - z_{ij}).
        \]
        \begin{enumerate}[label*=\arabic*.] 
            \item Set $\lambda_{i}^{\text{g}} = \lambda_{i}$ returned from Algorithm~\ref{salgo:DPSB}.
        \end{enumerate}
        
    \item For $j = 1, \dots, J-1$, generate local $\text{DPSB}(\boldsymbol{\eta}_{j}, \boldsymbol{\theta}_{j}, \alpha_{j}, \beta_{j})$ using Algorithm~\ref{salgo:DPSB} using $\{(n_{ij}, E_{ij}, z_{ij}); i = 1, \dots, I\}$ as data. 
            \begin{enumerate}
                \item[2.1.j] For this $j$, set $\lambda_{ij}^{l} = \lambda_{i}$  returned from Algorithm~\ref{salgo:DPSB}.
             \end{enumerate}
   
    \item Generate ``local'' allocations:
        \[
         z_{ij} \mid n_{ij}, E_{ij}, z_{ij},  \lambda_{ij}^l, \lambda_i^g \sim \text{Bernoulli}(p_{ij}),
        \]
        where
        \[
        p_{ij} = \frac{\pi f_{\text{Poisson}} (n_{ij} \mid \lambda_{ij}^{l} E_{ij})} {\pi f_{\text{Poisson}} (n_{ij} \mid  \lambda_{ij}^{l} E_{ij}) + (1 - \pi) f_{\text{Poisson}} (n_{ij} \mid \lambda_{i}^{g} E_{ij})}.
        \]

    \item Define $\lambda_{ij} = z_{ij} \lambda_{ij}^{l} + (1-z_{ij})\lambda_{i}^{g}$ for all $i = 1, \dots, I$ and $j = 1, \dots, (J-1)$.

    \item Generate $\pi$ using:
        \[
        \pi \mid z_{ij} \sim \text{Beta} \left(a_\pi + \sum_{i=1}^{I} \sum_{j=1}^{J-1} z_{ij}, b_\pi + \sum_{i=1}^{I} \sum_{j=1}^{J-1} (1 - z_{ij}) \right),
        \]

        \textbf{Notes:} $a_\pi$ = 1, $b_\pi$  = 1.
    
    \item Generate $\lambda_{1J}, \dots, \lambda_{IJ}$ from:
        \[
        \lambda_{iJ} \mid \tau, n_{iJ}, E_{iJ} \sim \text{Gamma}(\text{shape} = n_{iJ} + \tau, \text{rate} = E_{iJ} + \tau).
        \]

    \item Generate $\beta_{J}$ from:
        \[
        \pi(\beta_{J} \mid \lambda_{1J}, \dots, \lambda_{IJ}) \propto \prod_{i=1}^{I} f_{\text{Gamma}} (\lambda_{ij} \mid \text{shape} = \tau, \text{rate} = \tau) \pi_{\tau} (\tau).
        \]
    \end{enumerate}
    
\end{algorithm}

\begin{algorithm}[H]

\caption{One Iteration of DP ZIP local-global Model Given data $(n_{ij}, E_{ij}, z_{ij})$}
\label{salgo:zip}
    
    \begin{enumerate}
    \item Generate the ZI indicator $y_{ij}$  for $i = 1, \dots, I $, $j = 1, \dots, J$, where:
        \[
        y_{ij} \mid \omega_{j}, \lambda_{ij}, n_{ij}, E_{ij} \sim \text{Bernoulli}(p_{ij}^{\text{ZI}}),
        \]
    where
        \[
        p_{ij}^{\text{ZI}} =
        \begin{cases}
        \frac{\omega_{j}}{\omega_{j} + (1 - \omega_{j}) \exp(-\lambda_{ij} E_{ij})}, & \text{if } n_{ij} = 0, \\
        0, & \text{if } n_{ij} \geq 1.
        \end{cases}
        \]
    
    \item Generate $\omega_{1}, \dots, \omega_{J}$ independently with:
        \[
        \omega_{j} \mid y_{1j} \dots y_{Ij} \sim \text{Beta} \left( 1 + \sum_{i=1}^{I} y_{ij}, 1 + \sum_{i=1}^{I} (1 - y_{ij}) \right).
        \]

    \item Run Algorithm~\ref{salgo:poisson} with all $(n_{ij}, E_{ij})$ such that $y_{ij} = 0$.

    \end{enumerate}
\end{algorithm}

\subsection{Initial value generation for the MCMC algorithms}
\label{ssec:MCMC_initial}
Given the observed data $n_{ij}$ and $E_{ij}$ for $i = 1, \dots, I$, $j = 1, \dots, J$, and a prespecified number $K$ of DP mixture components for finite approximation. The default value for $K$ in our software is set to $\max(\log(IJ), 10)$, a setting that is generally sufficient for most cases. For the simulation study in Section~\ref{sec:simulation}, we used $K=10$, which satisfies the condition $K > \log(1492 \times 7)$. In the data analysis presented in Section~\ref{sec:data analysis}, we chose a more conservative value of $K=20$ to allow for a larger number of atoms in the DP, facilitating a more cautious interpretation of the results. We initialize the number of nonempty components $K_{+}$ to the smallest integer greater than $\min\{K/2, I-1\}$. The local-global indicators $z_{ij}$ are randomly initialized by independently sampling each cell $(i, j)$ from a $\text{Bernoulli}(0.5)$ distribution. The DP parameters for each local and global DP process are then initialized using the subset of data $(n_{ij}, E_{ij})$ assigned to that process, as determined by the indicators $z_{ij}$, for $i = 1, \dots, J - 1$.

For the initialization of parameters in the local DP prior for the $j$-th drug ($j = 1, \dots, J - 1$), we perform $k$-means clustering with the initialized $K_{+}$ number of clusters on the set $\{(n_{ij} + 0.5)/(E_{ij} + 0.5): z_{ij} = 1,\ i = 1, \dots, I\}$, and use the resulting cluster centers as initial values for the atoms $\theta_{hj}$, $h = 1, \dots, K_{+}$, of the non-empty clusters. The probability weights $\eta_{hj}$ for these non-empty clusters are initialized using the relative frequencies of the $k$-means clusters, and the clusters are then relabeled to ensure $\eta_{1j} \geq \eta_{2j} \geq \dots \geq \eta_{K_{+} j}$. These atoms $\theta_{hj}$ are then used for numerical maximum likelihood estimation of $\beta$ with the model $\theta_{1j}, \dots, \theta_{K_+, j} \sim \text{independent~Gamma}(\text{shape}=\beta, \text{rate}=\beta)$, and the resulting estimate $\hat \beta$ is used to initialize $\beta_{j}$. The atoms $\theta_{h'j}$ for the remaining empty clusters ($h' = K_{+} + 1, \dots, K$) are randomly drawn from $\text{Gamma}(\text{shape} = \beta_j, \text{rate} = \beta_j)$ using the initialized $\beta_j$. Their corresponding weights $\eta_{h'j}$ are initialized to $(\min_{h' \in {1, \dots, K_{+}}} \eta_{h'j})/10$, and all weights $\eta_{hj}$ for $h = 1, \dots, K$ are subsequently renormalized to ensure $\sum_{h=1}^K \eta_{hj} = 1$.  The local signal strengths $\lambda_{ij}^l$ are then initialized for each $i$ using the cluster center (atom) corresponding to the $k$-means cluster containing point $i$ if $z_{ij} = 1$, or by random sampling from ${\theta_{1j}, \dots, \theta_{Kj}}$ with probabilities ${\eta_{1j}, \dots, \eta_{Kj}}$. Finally, the concentration hyperparameter $\alpha_j$ is initialized by random sampling from a $\text{Gamma}(\text{shape} = 1, \text{rate} = 2)$ distribution. 

The parameters for the global DP prior are analogously initialized, but using the aggregated observed and expected counts $\{\tilde n_i = \sum_{j=1}^{J-1} n_{ij} (1-z_{ij}), \tilde E_i =  \sum_{j=1}^{J-1} E_{ij} (1-z_{ij})\}$ as the data with all AEs $i$ such that $\sum_{j=1}^{J-1} (1 - z_{ij}) > 0$. 

Finally, the reference drug $J$ shrinkage hyperparameter $\tau$ is initialized at $1$, while the zero-inflation parameters $\omega_j$ (if included in the model) are initialized each at $0.01$.

\subsection{Optimization method in pFDR calculation}
\label{ssec:pFDR_calculation}
A sequence of threshold values for $k$,  ranging from 1.1 to 3 in small increments, is established. For our computations we specified this grid to be \{1.1, 1.11, 1.12, 1.13, 1.14, 1.16, 1.17, 1.18, 1.19, 1.2, 1.25, 1.28, 1.31, 1.33, 1.36, 1.39, 1.42, 1.44, 1.47, 1.5, 1.6, 1.76, 1.91, 2.07, 2.22, 2.38, 2.53, 2.69, 2.84, 3\}. Similarly, a set of distinct small quantile levels for $p$ is chosen  (\{0.01, 0.015, 0.02, 0.025, 0.03, 0.035, 0.04, 0.045, 0.05, 0.055, 0.06, 0.065, 0.07, 0.075, 0.08, 0.085, 0.09, 0.095, 0.1\} in our computations). These define a range of possible combinations $(p,k)$ to be considered.

For every quantile level in the $p$-grid, we obtain a matrix of test statistics $T_{ij}(p)$ based on the $p$-th posterior quantiles for $\lambda_{ij}$ computed from their MCMC draws. Subsequently, we compute FDR$(p, k)$ and FNR$(p, k)$ estimates using the formulas presented in the main text (Section~\ref{sec:signal-detect}) for all possible $p$ and $k$ defined by the grid. From the set of all $(p,k)$ pairs, we then retain only those pairs with computed FDR $\leq \alpha$ for a preset threshold $\alpha$ ($=0.05$ in our analyses). Finally, among the retained $(p, k)$, we identify the pair that minimizes FNR. If multiple $(p, k)$ yield the same minimum FNR, we select the pair with the smallest $p$ and the largest $k$. 

\subsection{JAGS code for DP Hu et al. model}
\label{ssec:JAGS code}

\begin{lstlisting}[language=S, caption=JAGS exmpale][H]
    # Likelihood
    for (i in 1:I) {
      for (j in 1:J) {
        n[i, j] ~ dpois(E[i, j] * lambda[i, j])
      }
    }

    # Prior
    for (j in 1:J) {
      alpha[j] ~ dunif(0.2, 10)
      a[j] ~ dunif(0, 1)

      v[j, 1] ~ dbeta(1, alpha[j]); T(0.00001,0.99999) # to ensure in (0, 1)
      w[j, 1] <- v[j, 1]

      for (h in 2:K) {
        v[j, h] ~ dbeta(1, alpha[j]); T(0.00001,0.99999) # to ensure in (0, 1)
        w[j, h] <- v[j, h] * prod( (1 - c(v[j, 1:(h-1)])) )
      }

      for (h in 1:K) {
        theta[j, h] ~ dgamma(a[j], a[j]); T(0.000001, ) # to ensure > 0
      }

      for (i in 1:I) {
        ind[i, j] ~ dcat(c(w[j,]))
        lambda[i, j] <- theta[j, ind[i, j]]
      }
    }
\end{lstlisting}

\subsection{Example statin dataset}
\label{ssec:statin1491}
We present the first 30 rows of the reference SRS contingency table for statin drug in Table~\ref{stb:statin-data}. It is used for data generation in Section~\ref{sec:simulation} and data analysis in Section~\ref{sec:data analysis}.

\begin{table}[H]

\scriptsize
\caption{The first 30 rows of the SRS contingency table summarized from the FDA FEARS database (Q3 2014-Q4 2020) which catalogs a total of 63{,}976{,}610 reported cases. There are 1490 preferred terms in the full dataset.}
\label{stb:statin-data}
\begin{tabular}{|llllllll|}
\hline
Adverse Event                                & Atorvastatin & Fluvastatin & Lovastatin & Pravastatin & Rosuvastatin & Simvastatin & Other  \\ \hline
Abasia                                       & 117          & 0           & 7          & 9           & 84           & 45          & 18046  \\
Abdominal Adhesions                          & 10           & 0           & 0          & 0           & 3            & 2           & 3831   \\
Abdominal Discomfort                         & 501          & 33          & 14         & 92          & 497          & 217         & 214807 \\
Abdominal Distension                         & 267          & 7           & 5          & 57          & 251          & 262         & 101261 \\
Abdominal Pain                               & 655          & 27          & 7          & 116         & 428          & 372         & 232708 \\
Abdominal Pain Lower                         & 60           & 1           & 1          & 2           & 52           & 22          & 22527  \\
Abdominal Pain Upper                         & 696          & 14          & 15         & 78          & 554          & 266         & 197925 \\
Abdominal Rigidity                           & 12           & 0           & 0          & 1           & 1            & 12          & 2919   \\
Abdominal Symptom                            & 7            & 0           & 0          & 0           & 2            & 2           & 718    \\
Abdominal Tenderness                         & 20           & 0           & 0          & 0           & 5            & 10          & 5783   \\
Abnormal Behaviour                           & 72           & 2           & 5          & 9           & 23           & 77          & 35802  \\
Abnormal Faeces                              & 23           & 0           & 0          & 1           & 18           & 8           & 7982   \\
Abnormal Loss Of Weight                      & 89           & 0           & 1          & 3           & 15           & 79          & 9042   \\
Abnormal Sensation In Eye                    & 4            & 0           & 0          & 0           & 3            & 0           & 2571   \\
Abortion Induced                             & 10           & 0           & 0          & 3           & 1            & 55          & 6668   \\
Abortion Spontaneous                         & 32           & 2           & 0          & 5           & 12           & 4           & 28245  \\
Abscess Limb                                 & 4            & 0           & 0          & 0           & 4            & 0           & 5258   \\
Accident                                     & 24           & 0           & 0          & 9           & 28           & 6           & 12034  \\
Accident At Work                             & 0            & 0           & 0          & 0           & 14           & 2           & 2087   \\
Accidental Exposure To Product               & 13           & 2           & 1          & 3           & 12           & 7           & 46642  \\
Accidental Overdose                          & 56           & 0           & 0          & 11          & 7            & 36          & 42543  \\
Accidental Underdose                         & 4            & 0           & 0          & 0           & 2            & 0           & 7423   \\
Acidosis                                     & 55           & 3           & 0          & 4           & 4            & 22          & 8117   \\
Acne                                         & 28           & 0           & 1          & 10          & 35           & 22          & 105646 \\
Activities Of Daily Living Impaired          & 135          & 1           & 3          & 15          & 82           & 200         & 14483  \\
Acute Coronary Syndrome                      & 105          & 1           & 0          & 1           & 66           & 63          & 9259   \\
Acute Generalised Exanthematous   Pustulosis & 30           & 0           & 0          & 0           & 11           & 3           & 8161   \\
Acute Hepatic Failure                        & 107          & 0           & 10         & 1           & 18           & 66          & 14260  \\ \hline
\end{tabular}
\end{table}

We also present the full dataset with a smaller setting involving a subset of Table~\ref{stb:statin-data} with only $I=46$ relevant AEs as considered in prior work{\cite{chakraborty2022use, huang2011likelihood}}. It is used for data generation in the simulation of Section~\ref{ssec:simu_statin46}.

\begin{table}[H]

\scriptsize
\caption{The SRS contingency table summarized from the FDA FEARS database (Q3 2014-Q4 2020) which catalogs a total of 63{,}976{,}610 reported cases. There are 46 preferred terms in the full dataset.}
\label{stb:statin-data46}
\begin{tabular}{|llllllll|}
\hline
Adverse Event                                      & Atorvastatin & Fluvastatin & Lovastatin & Pravastatin & Rosuvastatin & Simvastatin & Other    \\ \hline
Acute Kidney Injury                       & 1353         & 42          & 7          & 154         & 689          & 823         & 355651   \\
Anuria                                    & 71           & 0           & 0          & 2           & 43           & 62          & 10403    \\
Blood Calcium Decreased                   & 14           & 2           & 0          & 0           & 110          & 17          & 15918    \\
Blood Creatine Phosphokinase Abnormal     & 34           & 0           & 0          & 0           & 8            & 11          & 261      \\
Blood Creatine Phosphokinase Increased    & 1175         & 125         & 32         & 200         & 562          & 768         & 23805    \\
Blood Creatine Phosphokinase Mm Increased & 2            & 0           & 0          & 0           & 9            & 0           & 14       \\
Blood Creatinine Abnormal                 & 27           & 0           & 0          & 0           & 5            & 3           & 3385     \\
Blood Creatinine Increased                & 227          & 10          & 0          & 17          & 210          & 97          & 74742    \\
Chromaturia                               & 340          & 10          & 6          & 33          & 174          & 114         & 19294    \\
Chronic Kidney Disease                    & 152          & 16          & 2          & 19          & 177          & 37          & 339179   \\
Compartment Syndrome                      & 53           & 0           & 0          & 1           & 12           & 12          & 2644     \\
Creatinine Renal Clearance Decreased      & 6            & 0           & 0          & 2           & 124          & 6           & 7768     \\
Diaphragm Muscle Weakness                 & 14           & 0           & 0          & 0           & 8            & 1           & 94       \\
Electromyogram Abnormal                   & 2            & 0           & 0          & 2           & 0            & 0           & 132      \\
End Stage Renal Disease                   & 30           & 0           & 0          & 0           & 19           & 6           & 97553    \\
Glomerular Filtration Rate Abnormal       & 8            & 0           & 0          & 0           & 1            & 0           & 1069     \\
Glomerular Filtration Rate Decreased      & 59           & 1           & 0          & 8           & 39           & 29          & 13190    \\
Hypercreatininaemia                       & 0            & 0           & 0          & 0           & 8            & 0           & 648      \\
Hypocalcaemia                             & 36           & 0           & 0          & 16          & 8            & 18          & 23102    \\
Muscle Disorder                           & 291          & 2           & 7          & 21          & 191          & 87          & 7329     \\
Muscle Enzyme Increased                   & 48           & 1           & 0          & 0           & 13           & 9           & 410      \\
Muscle Fatigue                            & 85           & 0           & 2          & 16          & 30           & 39          & 4257     \\
Muscle Haemorrhage                        & 24           & 0           & 0          & 5           & 13           & 4           & 3806     \\
Muscle Necrosis                           & 68           & 2           & 0          & 1           & 10           & 20          & 662      \\
Muscle Rupture                            & 181          & 25          & 0          & 61          & 36           & 120         & 3219     \\
Muscular Weakness                         & 1857         & 45          & 31         & 152         & 808          & 859         & 111003   \\
Musculoskeletal Discomfort                & 137          & 18          & 15         & 25          & 187          & 93          & 19931    \\
Musculoskeletal Disorder                  & 56           & 3           & 0          & 9           & 65           & 73          & 25881    \\
Musculoskeletal Pain                      & 420          & 3           & 2          & 38          & 324          & 228         & 82576    \\
Myalgia                                   & 5362         & 341         & 151        & 939         & 2757         & 3216        & 143819   \\
Myasthenic Syndrome                       & 1            & 0           & 9          & 0           & 3            & 6           & 643      \\
Myoglobin Blood Increased                 & 71           & 4           & 0          & 4           & 16           & 39          & 1003     \\
Myoglobin Blood Present                   & 2            & 0           & 0          & 0           & 0            & 0           & 8        \\
Myoglobin Urine Present                   & 0            & 0           & 0          & 0           & 1            & 2           & 70       \\
Myoglobinaemia                            & 15           & 0           & 0          & 0           & 0            & 0           & 62       \\
Myoglobinuria                             & 26           & 4           & 0          & 1           & 1            & 9           & 296      \\
Myopathy                                  & 849          & 64          & 45         & 145         & 377          & 544         & 6695     \\
Myopathy Toxic                            & 31           & 0           & 0          & 1           & 2            & 21          & 457      \\
Myositis                                  & 219          & 8           & 10         & 28          & 62           & 141         & 7482     \\
Necrotising Myositis                      & 279          & 0           & 0          & 2           & 10           & 52          & 278      \\
Oliguria                                  & 52           & 0           & 0          & 4           & 24           & 37          & 7590     \\
Renal Failure                             & 534          & 26          & 11         & 69          & 225          & 195         & 250710   \\
Renal Impairment                          & 390          & 52          & 11         & 37          & 161          & 169         & 103343   \\
Renal Tubular Necrosis                    & 40           & 0           & 0          & 12          & 10           & 24          & 12762    \\
Rhabdomyolysis                            & 2041         & 52          & 44         & 163         & 936          & 1376        & 31707    \\
Tendon Discomfort                         & 9            & 0           & 0          & 3           & 22           & 10          & 794      \\
Other Pt                                  & 180699       & 4886        & 2845       & 20296       & 113960       & 76068       & 61724222 \\ \hline
\end{tabular}
\end{table}

\section{ADDITIONAL SIMULATION RESULTS}

We present detailed results from our simulation experiments to compare among our local-global models (Poisson/ZIP), DP Hu et al. model, and a posterior quantile-based, FNR-optimized DP Hu et al. test method. Four evaluation metrics (FDR, sensitivity, average type I error, and F-score) are reported under the same simulation settings in Table~\ref{tb:simulation_scenarios} of the main test. Frequentist and Bayesian MSE are compared to further evaluate the loss of the estimation.

\subsection{Simulation results for DP-based models}
\label{ssec:simulation_additional}
We present the four evaluation metrics in Tables~\ref{stb:simu_DFR}, \ref{stb:Simu_Sensitivity}, \ref{stb:Simu_TIE}, and \ref{stb:Simu_F-score} to provide a clearer numerical comparison beyond Figure~\ref{fg:simulation}. Additionally, we further explored the effect of posterior quantile optimization on the DP Hu et al. in the process of FDR control and minimizing FNR. This allows for a more comprehensive understand the strategy of optimizing the posterior quantile of the test statistic and deterministic threshold simultaneously. 

We found that in the DP Hu et al. method, simultaneously optimizing the posterior quantile results in only a slight increase in FDR, which remains well-controlled within a very small level (around 0.01). In return, this optimization leads to an improvement in sensitivity, benefiting all cases. When comparing across all DP-based models, the sensitivity gain improvement brought by incorporating the local-global structure is more significant than from this optimization.

\begin{table}[H]
\centering
\tiny
\renewcommand{\arraystretch}{0.65}
\caption{Simulation of FDR for DP-based models} \label{stb:simu_DFR}
}   & 1.01      & 0.0000                                                                                & 0.0000                                                                            & 0.0000                           & 0.0050                                                                                               \\
                                                                                                   & 1.1       & 0.0000                                                                                & 0.0000                                                                            & 0.0000                           & 0.0074                                                                                               \\
                                                                                                   & 1.2       & 0.0102                                                                                & 0.0106                                                                            & 0.0061                           & 0.0153                                                                                               \\
                                                                                                   & 1.3       & 0.0100                                                                                & 0.0100                                                                            & 0.0051                           & 0.0120                                                                                               \\
                                                                                                   & 1.4       & 0.0076                                                                                & 0.0075                                                                            & 0.0044                           & 0.0115                                                                                               \\
                                                                                                   & 1.5       & 0.0059                                                                                & 0.0058                                                                            & 0.0044                           & 0.0109                                                                                               \\
                                                                                                   & 1.7       & 0.0043                                                                                & 0.0041                                                                            & 0.0039                           & 0.0094                                                                                               \\
                                                                                                   & 1.9       & 0.0031                                                                                & 0.0032                                                                            & 0.0036                           & 0.0082                                                                                               \\
                                                                                                   & 2         & 0.0031                                                                                & 0.0031                                                                            & 0.0033                           & 0.0078                                                                                               \\
                                                                                                   & 2.5       & 0.0024                                                                                & 0.0023                                                                            & 0.0026                           & 0.0063                                                                                               \\
                                                                                                   & 3         & 0.0015                                                                                & 0.0015                                                                            & 0.0026                           & 0.0059                                                                                               \\ \hline
\end{tabular}
\end{table}

\begin{table}[]

\centering
\tiny
\renewcommand{\arraystretch}{0.65}
\caption{Simulation of sensitivity for DP-based models} \label{stb:Simu_Sensitivity}
}   & 1.01      & 0.0000                                                                                & 0.0000                                                                            & 0.0000                           & 0.0000                                                                                               \\
                                                                                                   & 1.1       & 0.0039                                                                                & 0.0044                                                                            & 0.0004                           & 0.0007                                                                                               \\
                                                                                                   & 1.2       & 0.3061                                                                                & 0.3076                                                                            & 0.0492                           & 0.0593                                                                                               \\
                                                                                                   & 1.3       & 0.4517                                                                                & 0.4530                                                                            & 0.1213                           & 0.1356                                                                                               \\
                                                                                                   & 1.4       & 0.5458                                                                                & 0.5462                                                                            & 0.1876                           & 0.2041                                                                                               \\
                                                                                                   & 1.5       & 0.6058                                                                                & 0.6061                                                                            & 0.2461                           & 0.2644                                                                                               \\
                                                                                                   & 1.7       & 0.6804                                                                                & 0.6805                                                                            & 0.3397                           & 0.3588                                                                                               \\
                                                                                                   & 1.9       & 0.7149                                                                                & 0.7150                                                                            & 0.4131                           & 0.4330                                                                                               \\
                                                                                                   & 2         & 0.7275                                                                                & 0.7276                                                                            & 0.4441                           & 0.4636                                                                                               \\
                                                                                                   & 2.5       & 0.7669                                                                                & 0.7670                                                                            & 0.5584                           & 0.5761                                                                                               \\
                                                                                                   & 3         & 0.7960                                                                                & 0.7962                                                                            & 0.6299                           & 0.6462                                                                                               \\ \hline
\end{tabular}
\end{table}

\begin{table}[]

\centering
\tiny
\renewcommand{\arraystretch}{0.65}
\caption{Simulation of average type I error for DP-based models} \label{stb:Simu_TIE}
}   & 1.01      & 0.0000                                                                                & 0.0000                                                                            & 0.0000                           & 0.0000                                                                                               \\
                                                                                                   & 1.1       & 0.0000                                                                                & 0.0000                                                                            & 0.0000                           & 0.0000                                                                                               \\
                                                                                                   & 1.2       & 0.0000                                                                                & 0.0000                                                                            & 0.0000                           & 0.0000                                                                                               \\
                                                                                                   & 1.3       & 0.0001                                                                                & 0.0001                                                                            & 0.0000                           & 0.0000                                                                                               \\
                                                                                                   & 1.4       & 0.0001                                                                                & 0.0001                                                                            & 0.0000                           & 0.0000                                                                                               \\
                                                                                                   & 1.5       & 0.0000                                                                                & 0.0000                                                                            & 0.0000                           & 0.0000                                                                                               \\
                                                                                                   & 1.7       & 0.0000                                                                                & 0.0000                                                                            & 0.0000                           & 0.0000                                                                                               \\
                                                                                                   & 1.9       & 0.0000                                                                                & 0.0000                                                                            & 0.0000                           & 0.0000                                                                                               \\
                                                                                                   & 2         & 0.0000                                                                                & 0.0000                                                                            & 0.0000                           & 0.0000                                                                                               \\
                                                                                                   & 2.5       & 0.0000                                                                                & 0.0000                                                                            & 0.0000                           & 0.0000                                                                                               \\
                                                                                                   & 3         & 0.0000                                                                                & 0.0000                                                                            & 0.0000                           & 0.0000                                                                                               \\ \hline
\end{tabular}

\end{table}

\begin{table}[H]

\centering
\tiny
\renewcommand{\arraystretch}{0.65}
\caption{Simulation of F-score for DP-based models} \label{stb:Simu_F-score}
}   & 1.01      & 0.0000                                                                                & 0.0000                                                                            & 0.0000                           & 0.0000                                                                                               \\
                                                                                                   & 1.1       & 0.0068                                                                                & 0.0077                                                                            & 0.0008                           & 0.0014                                                                                               \\
                                                                                                   & 1.2       & 0.4644                                                                                & 0.4661                                                                            & 0.0934                           & 0.1113                                                                                               \\
                                                                                                   & 1.3       & 0.6186                                                                                & 0.6198                                                                            & 0.2158                           & 0.2380                                                                                               \\
                                                                                                   & 1.4       & 0.7031                                                                                & 0.7034                                                                            & 0.3151                           & 0.3378                                                                                               \\
                                                                                                   & 1.5       & 0.7519                                                                                & 0.7521                                                                            & 0.3942                           & 0.4167                                                                                               \\
                                                                                                   & 1.7       & 0.8079                                                                                & 0.8080                                                                            & 0.5062                           & 0.5262                                                                                               \\
                                                                                                   & 1.9       & 0.8324                                                                                & 0.8324                                                                            & 0.5836                           & 0.6023                                                                                               \\
                                                                                                   & 2         & 0.8409                                                                                & 0.8410                                                                            & 0.6140                           & 0.6315                                                                                               \\
                                                                                                   & 2.5       & 0.8670                                                                                & 0.8671                                                                            & 0.7156                           & 0.7290                                                                                               \\
                                                                                                   & 3         & 0.8857                                                                                & 0.8858                                                                            & 0.7718                           & 0.7829                                                                                               \\ \hline
\end{tabular}
\end{table}

\subsection{Simulation results for sensitivity analysis for grid configuration of $k$ in hypothesis testing}
\label{ssec:simulation_krange}

In our signal detection procedure, a sequence of threshold values for $k$ is constructed on a predefined grid. By default, we consider a grid ranging from $k = 1.1$ to $k = 3$, with small increments. This range is primarily motivated by the simulation design, where the maximum signal strength is set to be no larger than $3$, rendering this grid sufficient to capture all relevant signal levels under the data-generating mechanism.

To assess whether the default range for $k$ is restrictive, we conducted an additional sensitivity analysis under simulation Cases 2a. In this analysis, we adopted a highly conservative strategy by extending the upper bound of the grid to the maximum value observed among all MCMC draws of $\lambda$, effectively performing a global search over a substantially wider range of $k$. The resulting performance curves under this expanded grid are indistinguishable from those obtained using the default grid, across all evaluation metrics.

\begin{figure}[]
\centering
\includegraphics[width=\textwidth]{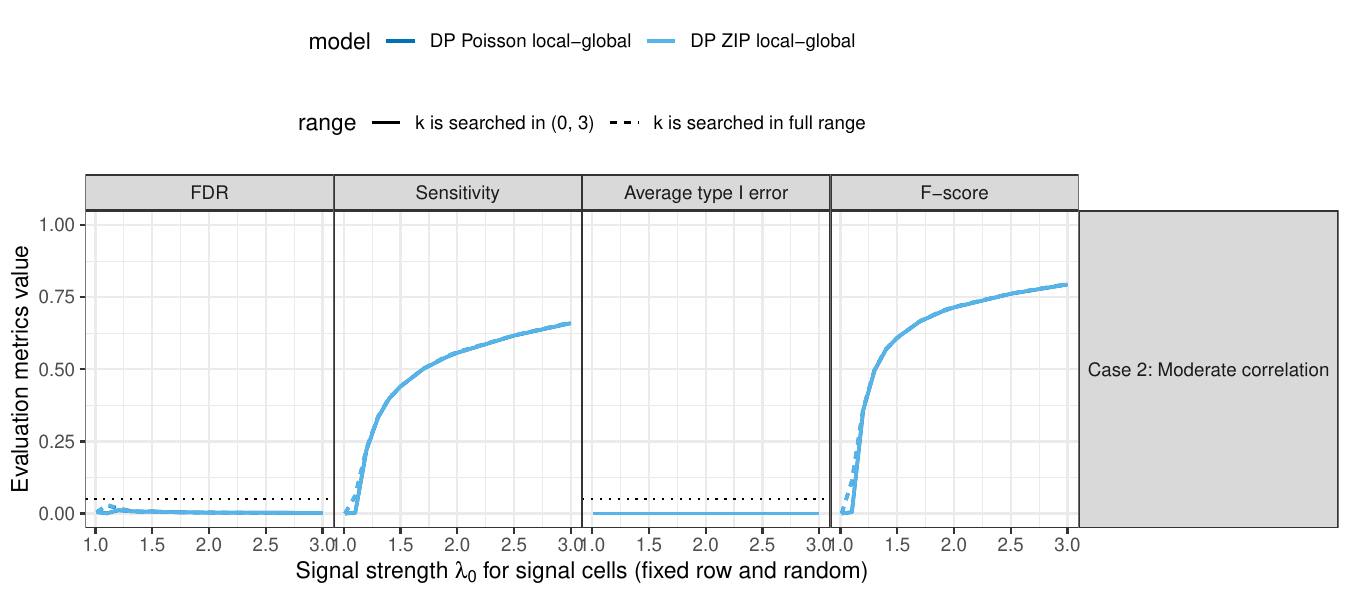} 

\label{fg:DP_krange}
\end{figure}

These results in Figure~\ref{fg:DP_krange} indicate that, for the considered simulation setting under Case 2a, the default grid range $(1.1, 3)$ is adequate and does not lead to any loss in detection performance. More importantly, the consistency between the two strategies suggests that the proposed method is robust to the choice of the search range for $k$. Consequently, in applications involving datasets with potentially stronger signals, adopting a more conservative, data-driven upper bound for $k$ remains a safe option and does not adversely affect performance.

\subsection{Simulation results for sensitivity analysis for hyperparameters}
\label{ssec:simulation_sensitivity}
We conducted a simulation study under Case 2a to evaluate the robustness of the proposed DP Poisson local-global model with respect to the choice of hyperparameters. The model structure was kept fixed throughout the simulations, and sensitivity analysis was performed by varying the hyperparameters governing the prior distributions of $\alpha$ and $\beta$, which control the strength and concentration of shrinkage in the local-global framework.

Specifically, the baseline model adopts $\psi_\alpha = 3$ and $\psi_\beta = 0.5$ as our default model. To assess sensitivity, we considered alternative hyperprior specifications that represent substantially different prior beliefs while remaining within reasonable ranges. For $\alpha$, we examined a more informative prior with $\psi_\alpha = 5$, placing approximately $99\%$ prior mass within $(0,1)$, as well as a weaker prior with $\psi_\alpha = 1$, placing approximately $64\%$ mass in $(0,1)$. For $\beta$, we considered $\psi_\beta = 1/7$, corresponding to a prior favoring values in $(0,1)$ with approximately $90\%$ prior mass in this interval, and $\psi_\beta = 6$, which places only $10\%$ prior mass in $(0,1)$ and thus favors more extreme shrinkage behavior.

These hyperprior configurations are summarized in Table~\ref{stab:hyperprior_sensitivity}. The comparison across these settings allows us to evaluate the stability of model performance under varying degrees of prior informativeness and shrinkage strength. 

\begin{table}[ht]
\centering
\caption{Hyperprior configurations used in the sensitivity analysis. The baseline setting corresponds to the default hyperprior choice in the proposed DP Poisson local--global model.}
\label{stab:hyperprior_sensitivity}
\begin{tabular}{lccp{10cm}}
\hline
Model setting & $\psi_\alpha$ & $\psi_\beta$ & Interpretation \\ 
\hline
Baseline & 3 & 0.5 & Default hyperprior specification in the proposed model \\

$\alpha$-strong & 5 & 0.5 & More informative prior on $\alpha$, placing $\approx 99\%$ mass in $(0,1)$ \\

$\alpha$-weak & 1 & 0.5 & Weaker prior on $\alpha$, placing $\approx 64\%$ mass in $(0,1)$ \\

$\beta$-favoring & 3 & $1/7$ & Prior favoring $\beta \in (0,1)$ with $\approx 90\%$ mass in $(0,1)$ \\

$\beta$-sparse & 3 & 6 & Prior discouraging $\beta \in (0,1)$, placing only $10\%$ mass in $(0,1)$ \\
\hline
\end{tabular}
\end{table}

\begin{figure}[]
\centering
\includegraphics[width=\textwidth]{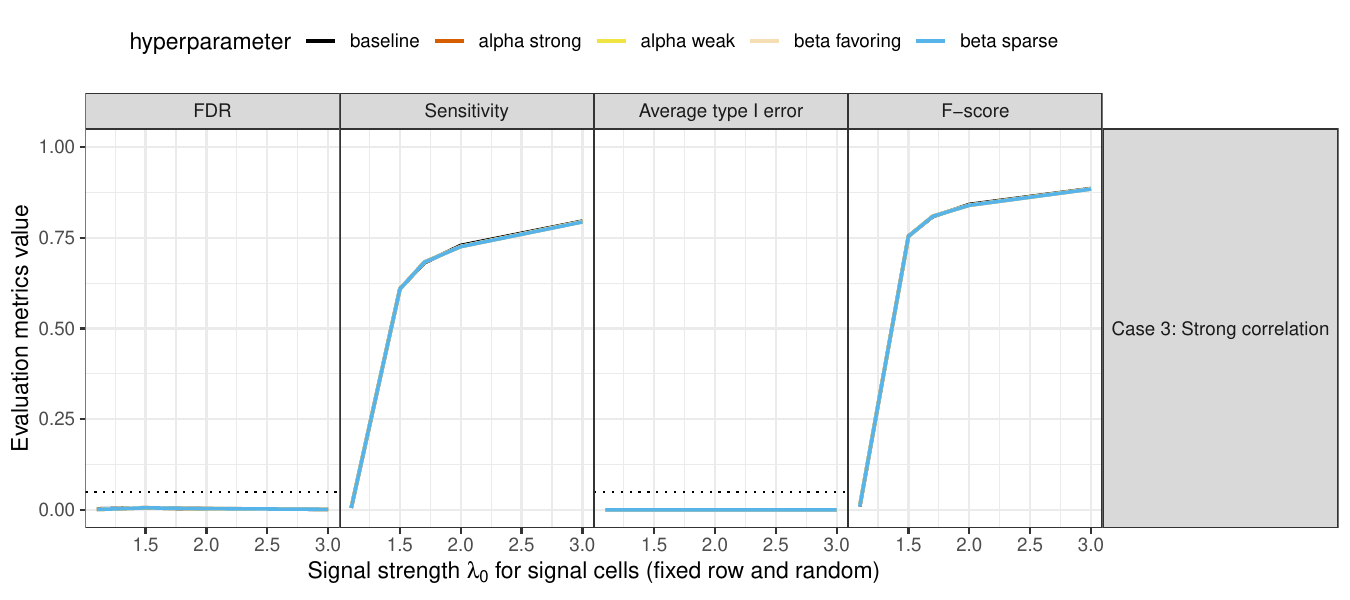} 

\caption{Simulation results under different hyperprior configurations for the DP Poisson local-global model. Curves corresponding to all hyperprior settings largely overlap across performance metrics, indicating robust model performance with respect to reasonable variations in hyperparameter choices.}
\label{fg:DP_sensitivity}
\end{figure}

As shown in Figure~\ref{fg:DP_sensitivity}, across all performance metrics, the curves corresponding to different hyperprior configurations largely overlap in the simulation results. This indicates that the proposed DP Poisson local-global model exhibits stable performance under a wide range of hyperparameter choices. In particular, varying the hyperpriors for $\alpha$ and $\beta$, from relatively informative to weakly informative specifications, does not materially affect model behavior at the considered data scale. These findings suggest that the proposed model is reasonably robust to reasonable variations in hyperprior assumptions. Moreover, the sensitivity analysis covers several representative hyperprior scenarios, which together are expected to reflect a broad range of practical data-generating settings.

\subsection{Simulation results under smaller setting}
\label{ssec:simu_statin46}

\begin{figure}[htpb]
\centering
\includegraphics[width=\linewidth]{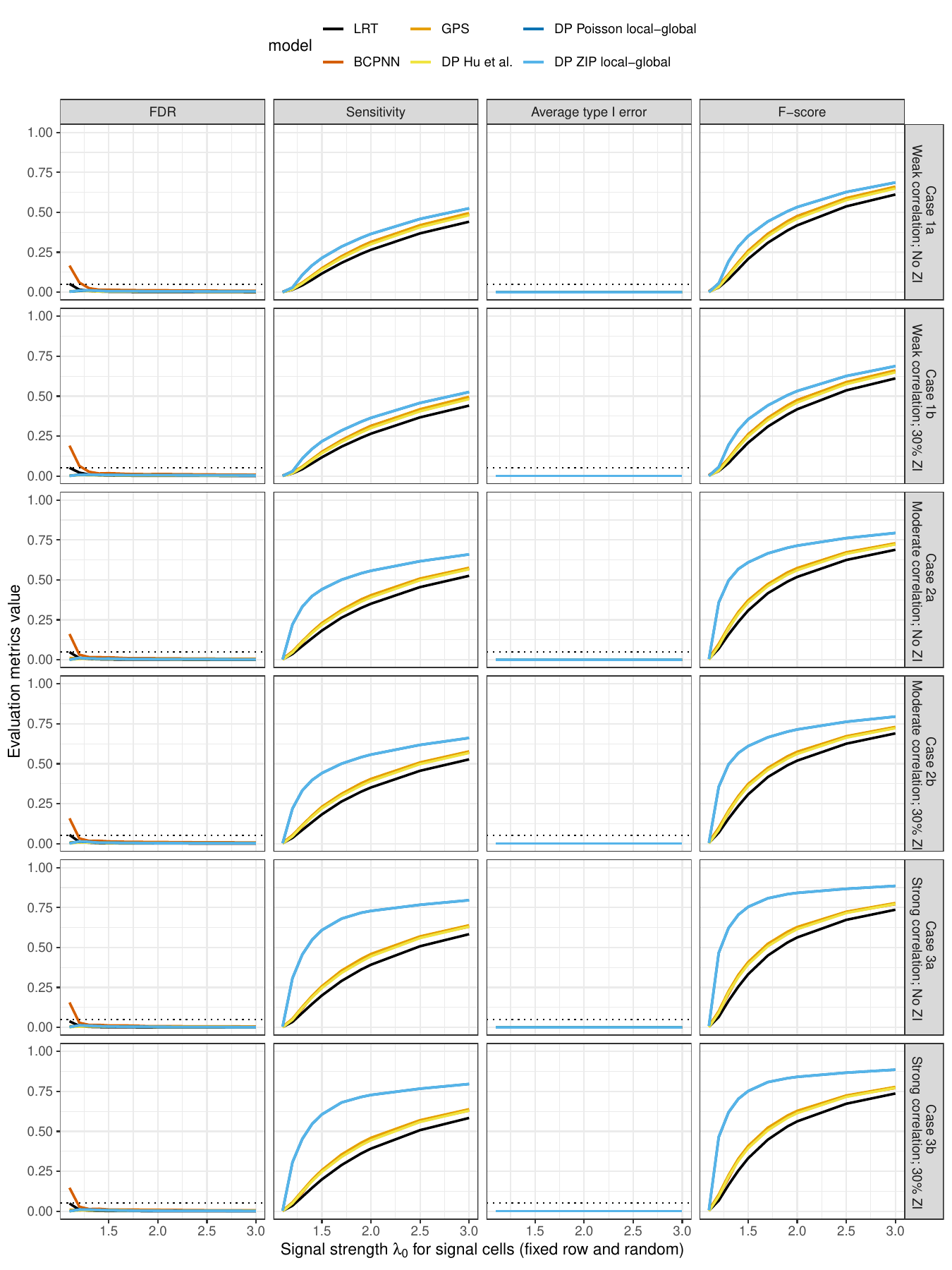} 
\caption{Simulation results: the horizontal axis shows the true signal strength $\lambda_0$ for signal cells, the vertical axis shows the value of evaluation metrics. Each column presents one metric and each row presents one simulation scenario.}
\label{sfg:simulation_statin46}
\end{figure}

We conduct the simulation based on Table~\ref{stb:statin-data46} under similar correlation strength setting in Table~\ref{tb:simulation_scenarios}. The result shown in Figure~\ref{sfg:simulation_statin46} has the same pattern as we observed in Section~\ref{sec:simulation} based on larger setting.

\end{document}